\documentclass[twocolumn,aps,prd,nofootinbib]{revtex4}
\usepackage{aas_macros}
\usepackage{amsmath,amssymb,graphicx,bm}
\usepackage{color}
\definecolor{darkgreen}{rgb}{0,0.6,0}

\newcommand\kms{\mbox{km~s$^{-1}$}}
\newcommand\ud{\mathrm{d}}
\newcommand\Msun{$\mathrm{M}_\odot$}

\newcommand{\PlaceFigCumulativeMass}{
\begin{figure}
\begin{center}
\includegraphics[width=\columnwidth]{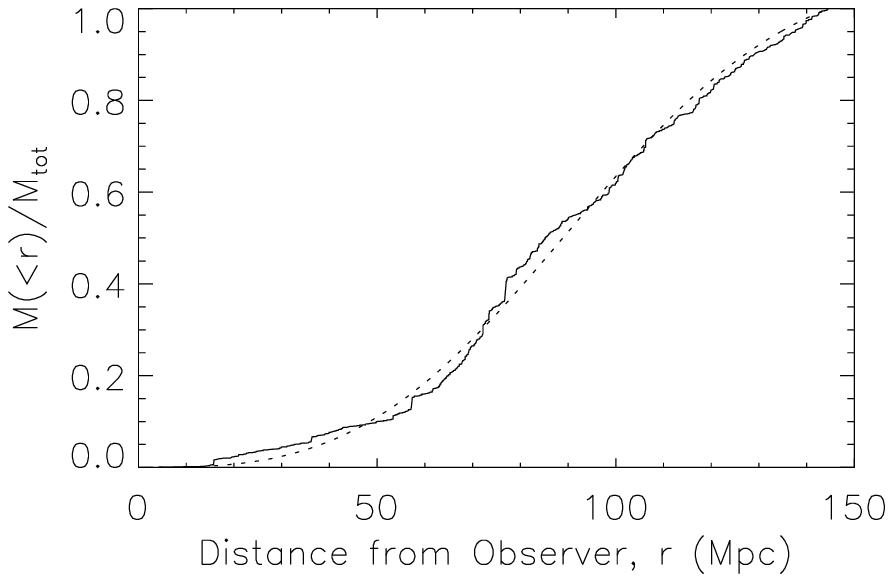} 
\end{center}
\caption{\label{Fig:CumulativeMass}
Mass in observed groups. The solid line shows the mass (as a fraction of the total mass), estimated dynamically, in observed groups of galaxies with 5 or more members inside a sphere of radius $r$, as a function of $r$. The dotted line shows the function of form (\ref{Eq:CumulativeMass}) obtained with a least-squares fit. The deviations from the fit for nearby distances are due to the overdensity in the Local Supercluster.
}
\end{figure}
}

\newcommand{\PlaceFigAngYukawa}{
\begin{figure*}
\begin{center}
\includegraphics[width=0.95\columnwidth]{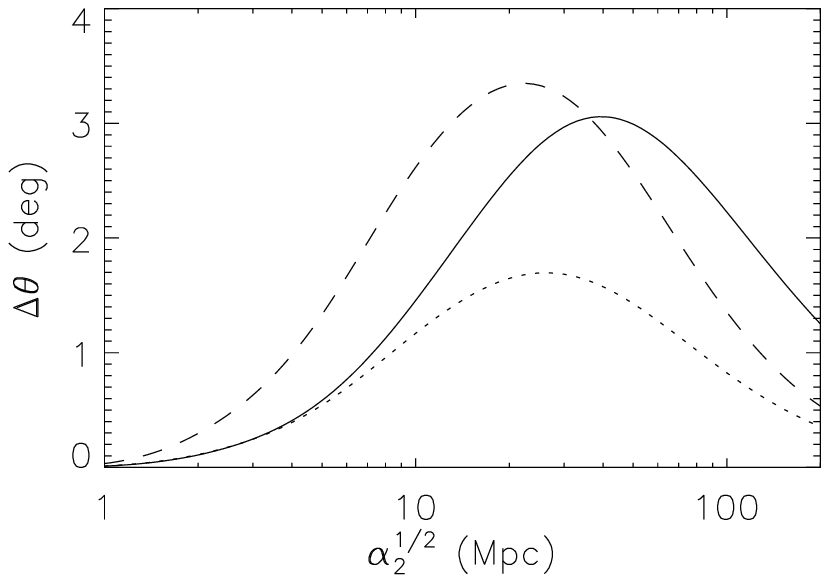}
\hspace{5mm}
\includegraphics[width=0.95\columnwidth]{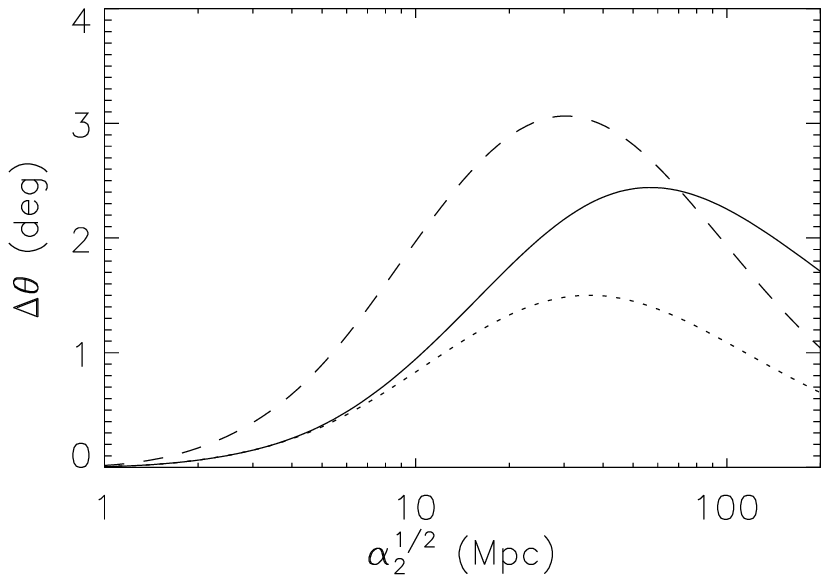}
\end{center}
\caption{\label{Fig:Ang_Yukawa}
Angular deviation of the LG peculiar velocity reconstructed under Yukawa-type modified gravity (or dark-interactions), from the direction of the standard GR reconstruction. The left panel shows the case $s=2$ and the right panel $s=4$, for three different values of the effective coupling $\alpha_1/\alpha_2$: 0.5 (solid line), 1.5 (dotted line), and 2.3 (dashed line).
}
\end{figure*}
}

\newcommand{\PlaceFigMagYukawa}{
\begin{figure*}
\begin{center}
\includegraphics[width=0.95\columnwidth]{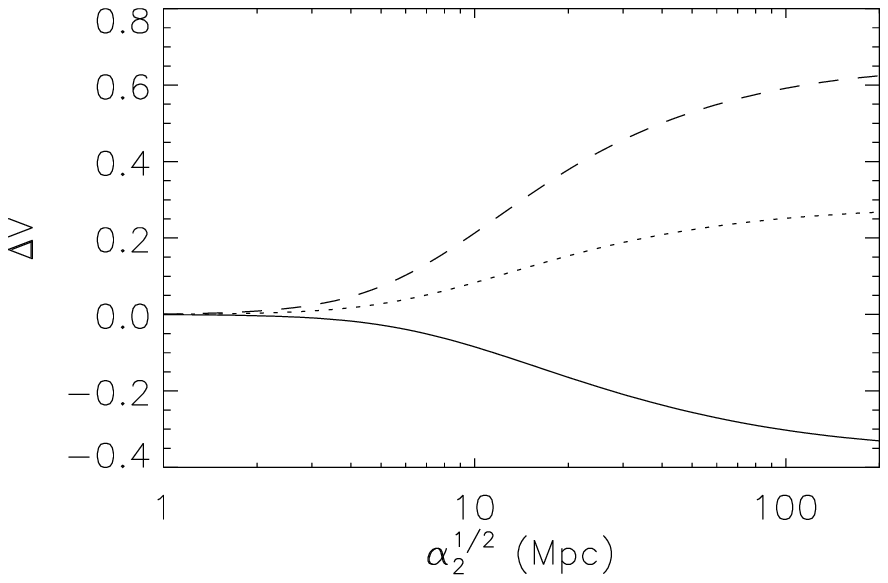}
\hspace{5mm}
\includegraphics[width=0.95\columnwidth]{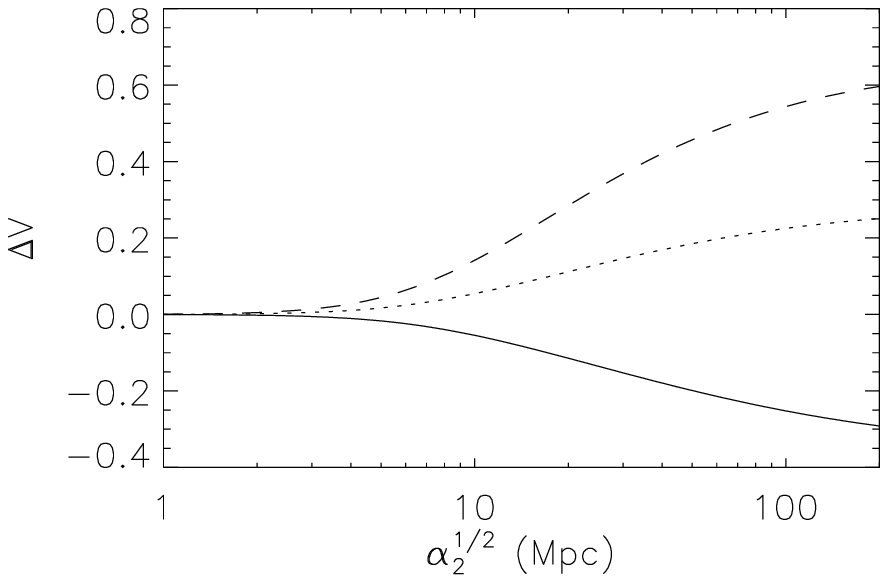}
\end{center}
\caption{\label{Fig:Mag_Yukawa}
Deviation in magnitude of the reconstructed LG peculiar velocity relative to the GR result, under Yukawa-type modified gravity (or dark-interactions). The figures show $\Delta v$ (eq.~\ref{Eq:DeltaV}) as a function of $\sqrt{\alpha_2}$. The left panel shows the case $s=2$ and the right panel $s=4$, for three different values of the effective coupling $\alpha_1/\alpha_2$: 0.5 (solid line), 1.5 (dotted line), and 2.3 (dashed line).
}
\end{figure*}
}

\newcommand{\PlaceFigConvergenceYukawa}{
\begin{figure*}
\begin{center}
\includegraphics[width=0.95\columnwidth]{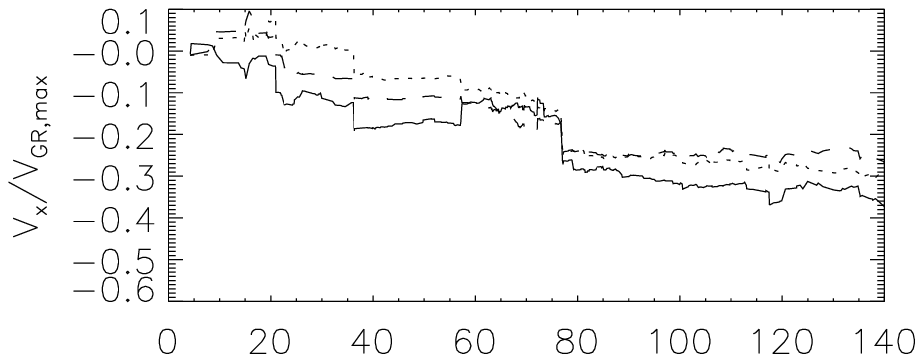}
\hspace{5mm}
\includegraphics[width=0.95\columnwidth]{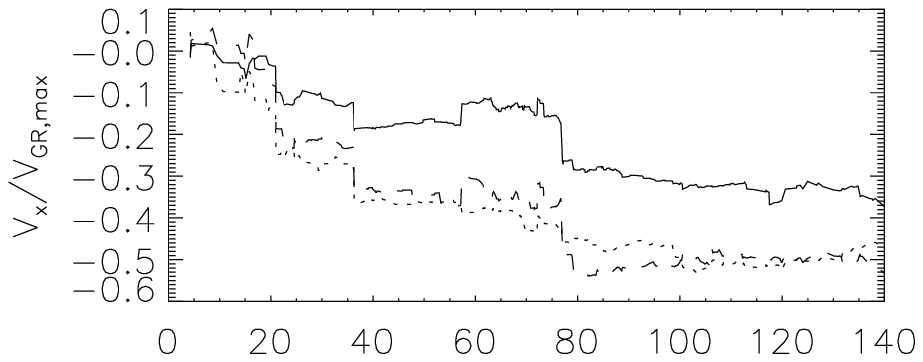}
\vspace{1mm}
\includegraphics[width=0.95\columnwidth]{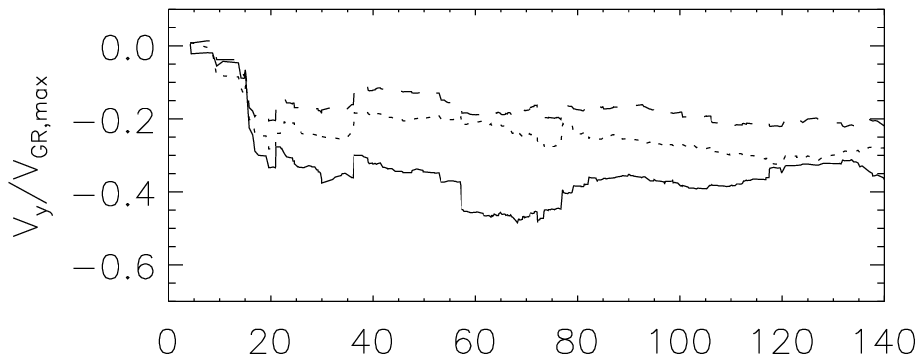}
\hspace{5mm}
\includegraphics[width=0.95\columnwidth]{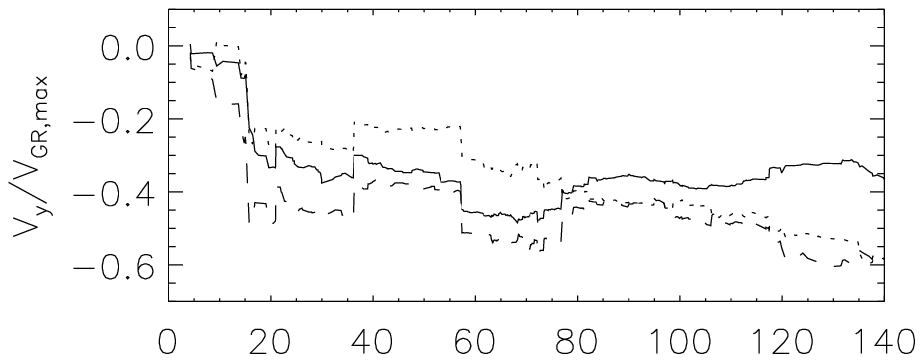}
\vspace{1mm}
\includegraphics[width=0.95\columnwidth]{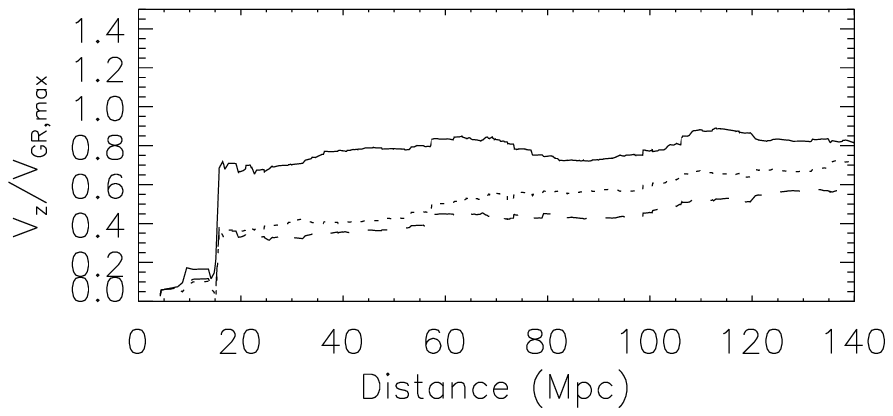}
\hspace{5mm}
\includegraphics[width=0.95\columnwidth]{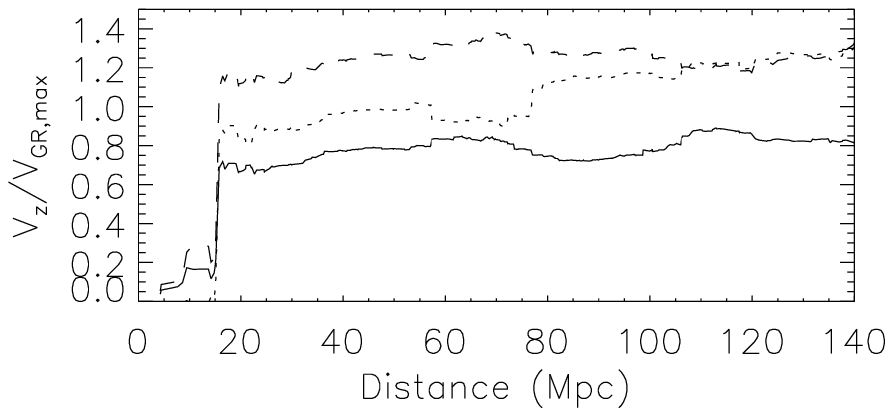}
\end{center}
\caption{\label{Fig:Convergence_Yukawa}
Components of the LG peculiar velocity as a function of the size of sphere-of-influence under Yukawa-type modified gravity. The graphs are constructed using $s=2$. The left panel shows the case $\alpha_1/\alpha_2=0.5$ and the right panel $\alpha_1/\alpha_2=2.3$. The graphs show the cartesian components of the LG peculiar velocity in galactic coordinates.
The vertical axis is normalized such that at the limiting distance of the sample (140 Mpc), the magnitude of the GR--case velocity is unity.
The solid line represents the GR reconstruction; the dotted and dashed lines represent the peculiar velocities reconstructed for two different scale lengths: $\sqrt{\alpha_2} = 30\,{\rm Mpc}$ and $100\,{\rm Mpc}$, respectively. Note that the vertical axes of the different cartesian components do not have the same scales.
}
\end{figure*}
}

\newcommand{\PlaceFigAngDegrav}{
\begin{figure}
\begin{center}
\includegraphics[width=\columnwidth]{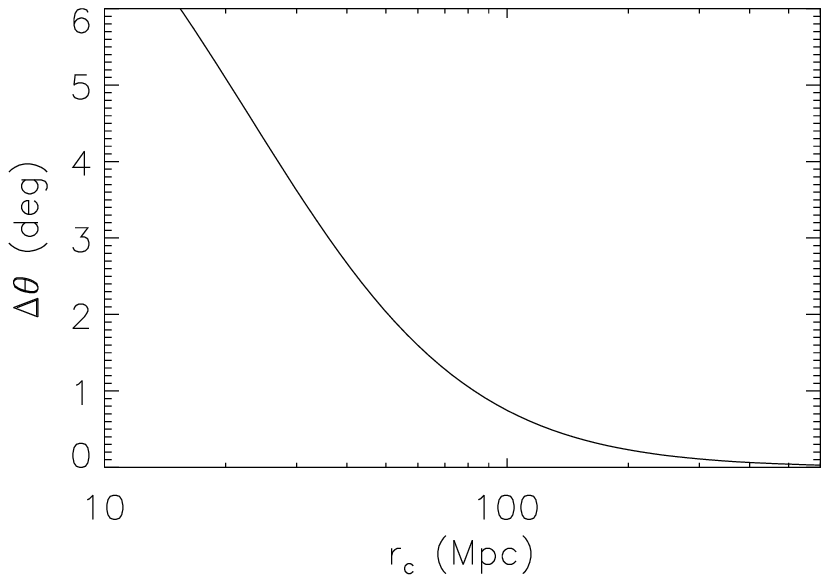}
\end{center}
\caption{\label{Fig:Ang_Degrav}
Angular deviation of the LG peculiar velocity reconstructed under degravitation, from the direction of the standard GR reconstruction
}
\end{figure}
}

\newcommand{\PlaceFigMagDegrav}{
\begin{figure}
\begin{center}
\includegraphics[width=\columnwidth]{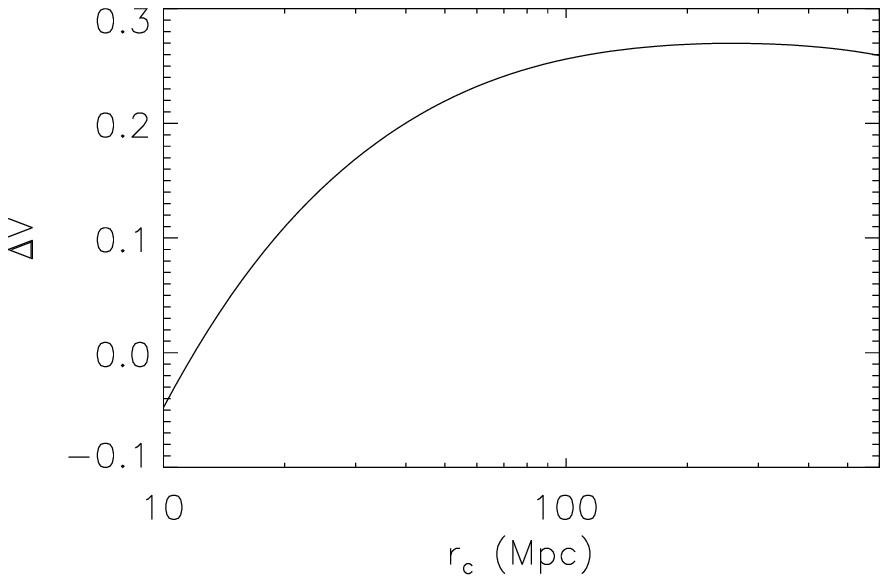}
\end{center}
\caption{\label{Fig:Mag_Degrav}
Deviation in magnitude of the reconstructed LG peculiar velocity for the higher-dimensional gravity case, relative to the GR result. The figures show $\Delta v$ (eq.~\ref{Eq:DeltaV}) as a function of $r_c$.
}
\end{figure}
}

\newcommand{\PlaceFigConvergenceDegrav}{
\begin{figure}
\begin{center}
\includegraphics[width=\columnwidth]{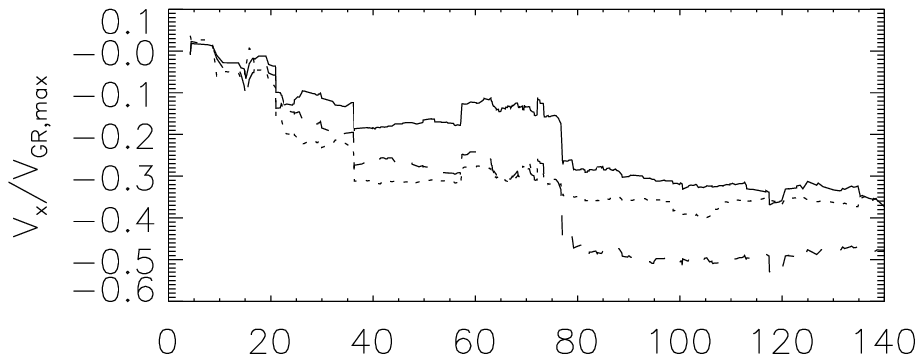}
\vspace{1mm}
\includegraphics[width=\columnwidth]{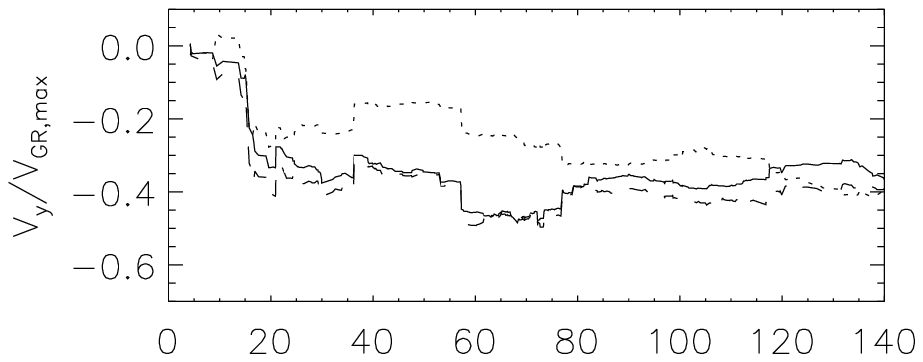}
\vspace{1mm}
\includegraphics[width=\columnwidth]{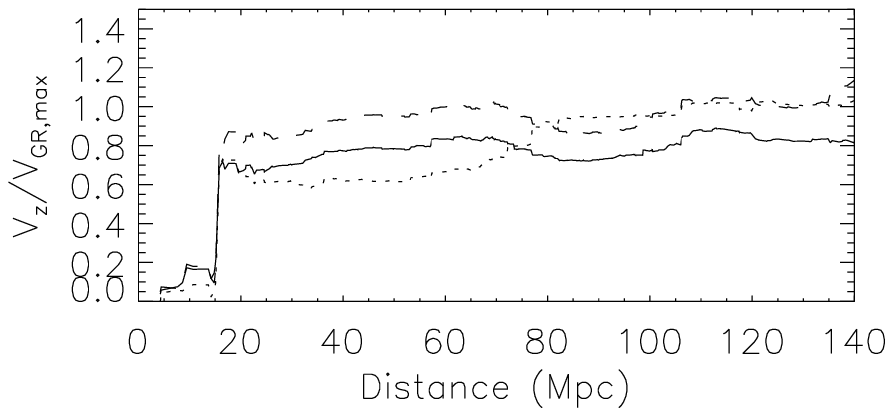}
\end{center}
\caption{\label{Fig:Convergence_Degrav}
Components of the LG peculiar velocity as a function of size of sphere-of-influence under degravitation formulation. 
The vertical axis is normalized such that at the limiting distance of the sample (140 Mpc), the magnitude of the GR--case velocity is unity.
The graphs show the cartesian components of the LG peculiar velocity in galactic coordinates. The solid line represents the GR reconstruction; the dotted and dashed lines represent the peculiar velocities reconstructed for two different scale lengths: $r_c$ = 20 Mpc and 300 Mpc, respectively. While $r_c=20$ Mpc is not physically meaningful, it is insightful to understand how a theory of this form could change the LG peculiar velocity. Note that the vertical axes of the different cartesian components do not have the same scales.
}
\end{figure}
}

\begin{document}

\title{Reconstructing the Peculiar Velocity of the Local Group with Modified Gravity and 2MASS}

\author{Aidan C Crook}
\email{acc@space.mit.edu}
\author{Alessandra Silvestri}
\author{Phillip Zukin}
\thanks{(authors contributed equally to this work)}
\affiliation{
Department of Physics, Massachusetts Institute of Technology, Cambridge, MA 02139, USA\\
}

\begin{abstract}
The peculiar velocity of the Local Group, reconstructed from inhomogeneities in the local density field, differs in direction and magnitude from the velocity inferred from the Cosmic Microwave Background dipole. We investigate whether generalized theories of gravity, which predict a modified growth of perturbations,  are able to alleviate this discrepancy.
We introduce a general formalism for calculating the real-space peculiar-velocity field for modified gravity and theories with interactions in the dark sector.
For different classes of theories --- scalar tensor and higher-dimensional gravity  --- we reconstruct the Local Group peculiar velocity using groups of galaxies identified in the 2MASS Redshift Survey.
We show that, for realistic parameters, modifications to General Relativity cannot account for the angular discrepancy between the reconstructed Local Group velocity and the dipole in the Cosmic Microwave Background.
\end{abstract}

\maketitle

\section{Introduction} \label{sec:Intro}
A little over a decade after the discovery of the Cosmic Microwave Background (CMB) \citep{Penzias:1965}, reports emerged of a dipole anisotropy in its spatial temperature distribution~\citep{Corey:1976,Fabbri:1980}. The origin of the dipole anisotropy is now widely accepted as a Doppler effect arising from the motion of the Sun with respect to the CMB rest frame. Measurements by the COsmic Background Explorer (COBE)~\cite{Smoot:1992,Bennett:1993}, confirmed more recently with WMAP, imply a Solar System velocity with respect to the CMB of $369.0 \pm 0.9$~\kms~\citep{Hinshaw:2009}. Efforts to infer the motion of the Sun with respect to the Local Group (hereafter, LG) centroid~\citep[e.g.][]{Yahil:1977, Courteau:1999} further imply that the LG has a net peculiar velocity in excess of $600~{\rm \kms}$~with respect to the rest frame of the CMB, in the direction $(l,b)\sim(276^{\circ},30^{\circ})$~\citep[e.g.][]{Maller:2003,Erdogdu:2006d}.

The large peculiar velocity of the LG is believed to have been gravitationally induced. A number of studies have attempted to make a comparison between the LG peculiar velocity inferred from the dipole in the CMB and the velocity reconstructed from inhomogeneities in the local density field \citep[e.g.,][]{Webster:1997, RowanRobinson:2000, Maller:2003, Erdogdu:2006d, Lavaux:2008}. The results vary significantly depending on both the source of the survey data as well as the reconstruction technique used. 

In reconstructing the velocity from local inhomogeneities, one needs to assume a bias factor to infer the total mass density from the observed one. For the usual assumption of a linear bias, the predicted magnitude of the reconstructed velocity  depends on the value of the bias, while its direction is insensitive to any change in the value of the bias; therefore a discrepancy in the angle of the reconstructed LG peculiar velocity is of more significance than a discrepancy in the magnitude.
The most recent effort by \citet{Lavaux:2008} predicts a LG velocity $\sim 50^\circ$ from the CMB direction, with an estimated error of $22^\circ$ (95\% confidence) from the reconstruction technique.

There are a number of possible reasons for this large discrepancy. Since the velocity is reconstructed from a survey of galaxies, it is  
natural to wonder whether the discrepancy arises because of the finite size of the survey. Including the error caused by a finite survey volume, as well as the error associated with the reconstruction method, the work of \citet{Lavaux:2008} suggests the dipole directions should agree to within 25$^\circ$ at 95\% confidence.
Indeed, there are a number of other possible explanations -- from reduced survey completeness in the zone of avoidance, to invalid assumptions in the estimation of mass, to the assumption of linear bias. Unfortunately, due to the limitations of the surveys and available data, it is difficult to avoid making assumptions of this kind. An analysis of each of these postulates is beyond the scope of this paper. We focus on the assumption that General Relativity (GR) is the correct theory of gravity. This assumption of GR is common to all previous reconstruction efforts. 

While GR has been tested to high accuracy on small scales, such as the solar system~\cite{Will}, the constraints coming from cosmological tests are weaker. In this paper we explore the effects of large-scale modifications of gravity on the reconstruction of the LG peculiar velocity obtained from an analysis of groups of galaxies in the 2MASS Redshift Survey~\citep{Huchra:2005r,Huchra:2009,Crook:2007,Crook:2008}.
We consider models of modified gravity and interacting dark energy that serve as alternatives to $\Lambda$CDM and that modify the growth of perturbations in a time- and scale-dependent manner.
Models of modified gravity that predict a scale-independent growth of structure cannot modify the direction of the LG peculiar velocity with respect to the GR prediction, since the peculiar velocity toward each attractor would simply be rescaled by a common factor in this case. However, a scale-dependent modification will change the weight of objects at different distances from the observer, and is therefore expected to rotate the direction. 

After generalizing the relationship between the peculiar velocity and density field to allow for modifications of GR and interactions in the dark sector, we study two classes of models
that introduce scale-dependent modifications. The first class of models we consider is inspired by scalar-tensor, e.g. $f(R)$, as well as coupled dark energy theories; generically, they introduce a Yukawa-type correction to the Newtonian potential. The second class is motivated by extra-dimensional theories~\cite{Dvali,Rham} and introduces a mass scale for the graviton.

The paper is organized as follows. In Sec. \ref{sec:PecVel} we generalize the relationship between the peculiar-velocity field and the density field in real space to models of modified gravity and models with interactions in the dark sector. We then specialize to the specific cases mentioned above. In Sec. \ref{sec:Reconstruction} we discuss our reasons for utilizing the 2MASS Groups sample and describe the method used to reconstruct the LG peculiar velocity from data in the 2MASS Redshift Survey. We discuss our findings in Sec. \ref{sec:Results}, and summarize our conclusions in Sec. \ref{sec:Summary}.


\section{Peculiar Velocities in Modified Gravity} \label{sec:PecVel}
In the linear regime, assuming zero pressure, the continuity equation reads
\begin{equation}\label{continuity}
\bar{\nabla}\cdot\bar{v}=-a\frac{d\delta}{dt}
\end{equation} 
\noindent where $\bar{v}$ is the peculiar velocity field, $a$ is the scale factor, $\delta \equiv \delta\rho/\bar{\rho}$ is the matter density contrast, $\bar{\rho}$ is the background density, $t$ is the cosmic time and the spatial derivatives are taken w.r.t. comoving coordinates. For modified gravity, eq.~(\ref{continuity}) holds generically since the energy-momentum conservation equations are left unchanged. For models in which there is an additional interaction in the dark sector, like coupled quintessence models~\cite{Bean:2001ys,Amendola:2003wa}, the conservation equations are modified, since the energy-momentum tensor for dark matter is not separately conserved. However, in the sub-horizon limit, eq.~(\ref{continuity}) holds approximately for dark matter~\cite{Amendola:2003wa}, and we can use it for our reconstruction algorithm. 

Assuming vorticity is negligible on the scales we consider~\cite{Pichon}, we can rewrite the peculiar velocity in terms of the velocity potential $\bar{v}\equiv\bar{\nabla}\phi_v$. Plugging this into the continuity equation (\ref{continuity}), and transforming to Fourier space, we obtain:

\begin{equation}\label{vel_pot}
\tilde{\phi}_v=\frac{aH}{k^2}f(k,a)\delta_k(a)\,,
\end{equation}

\noindent where $H\equiv d\ln a/dt$, $k$ is the comoving wave vector, and $f(k,a)\equiv d\ln\delta/d\ln a$ is the {\it growth factor}.  

We can solve for the growth factor by combining the Euler equation and the continuity equation (\ref{continuity}) to find the equation for the growth of structure:
\begin{equation}\label{eq-delta}
\frac{d}{dt}\left(a^2\frac{d\delta_k}{dt}\right)=-k^2\tilde{\Phi}\,,
\end{equation}

\noindent where $\tilde{\Phi}$ is the (Fourier transformed) time-time component of the perturbed metric in Newtonian gauge. In models with a coupling in the dark sector, the r.h.s. of eq.~(\ref{eq-delta}) contains a contribution from the coupling term in the Euler equation. As a result, dark matter particles respond to an effective potential; however, this effect can be absorbed in a modification of the Poisson equation which we introduce below. 

We shall now substitute for $\tilde{\Phi}$ in terms of the density contrast. This can be done via the Einstein's equations, which in general will be modified. We introduce a generic function of time and scale, $\mu_k(a)$, to describe how modified theories change the relationship between $\tilde{\Phi}$ and $\delta_k$ on sub-horizon scales:
\begin{equation}\label{Poisson}
-k^2\tilde{\Phi}= 4\pi Ga^2\bar{\rho}(a) \mu_k(a)\delta_k\,.
\end{equation}

Combining equations (\ref{eq-delta}) and (\ref{Poisson}), and rewriting the resulting equation in terms of the growth factor $f$ and $\ln a$, we obtain a nonlinear differential equation for the growth factor:
\begin{equation}\label{growth}
\frac{df}{d\ln a}+f^2+\left(2+\frac{d\ln H}{d\ln a}\right)f=\frac{3}{2}\Omega_m(a)\mu_k(a)\,,
\end{equation}
 where $\Omega_m(a)\equiv 8\pi G\bar{\rho}_m(a)/3H^2(a)$.
After solving eq.~(\ref{growth}) for $f(a,k)$, we obtain the velocity potential in real space via the convolution theorem:
\begin{equation}\label{vel_pot_real_1}
\phi_v(\bar{x},a)=\frac{aH}{4\pi\bar{\rho}(a)}\int d^3x'\delta\rho(a,\bar{x}'){\mathcal F}(\bar{x}-\bar{x}',a)\,,
\end{equation}
where $\bar{x}$ is the comoving distance and ${\mathcal F}$ is the inverse Fourier transform of $4\pi f(k,a)/k^2$. Finally, by taking the gradient of~(\ref{vel_pot_real_1}) w.r.t. $x'$ we obtain the velocity in real space:

\begin{equation}\label{vel__real_1}
\bar{v}(\bar{x},a)=\frac{aH}{4\pi\bar{\rho}(a)}\int d^3x'\delta\rho(a,\bar{x}')\bar{{\mathcal F}}_v(\bar{x}-\bar{x}',a),
\end{equation}
where $\bar{{\mathcal F}}_v\equiv\bar{\nabla}{\mathcal F}$.

In our reconstruction algorithm we first parametrize different theories via the function $\mu_k(a)$, then solve eq.~(\ref{growth}) for $f(k,a=1)$ subject to the initial condition $f(k,10^{-3})=1$, and perform the inverse Fourier transform of $f(k,a=1)/k^2$ to obtain ${\mathcal F}(\bar{x}-\bar{x}',a=1)$ and $\bar{\mathcal F}_v$. The initial condition for the growth factor is taken to correspond to the GR solution during matter domination. 
In other words, the theory must reduce to GR at early times. The reconstruction methods based on GR commonly use the approximation $f(a)=\Omega_m(a)^{6/11}$ \cite[e.g.][]{Wang:1998}; in our analysis we compare the predictions of modified models of gravity to the GR result obtained by setting $\mu_k(a)=1$.

While the equation for the growth factor~(\ref{growth}) can be solved numerically, 
we can also proceed analytically with some approximations.  
Although the analytical method has limited applicability, as we will discuss, it is very useful in understanding the effects of modified gravity. 
Moreover it works well in the relevant range of parameters.  In the remainder of this section we present the analytical results and then describe the two classes of modified gravity models that we consider in our analysis.

The equation for the growth factor can be solved by quadratures if one neglects terms of $O((f-1)^2)$. The approximate solution is given by~\cite{Linder}:

\begin{equation}\label{growth_factor}
f(k,a)\approx\frac{1}{a^4H}\int^a_0da'\,a'^3H(a')\left(1+\frac{3}{2}\Omega_m(a')\mu_k(a')\right)\,,
\end{equation}

Expression~(\ref{growth_factor}) is a good approximation when $f$ does not depart significantly from unity throughout the expansion history. It systematically overestimates the actual growth factor.  
For $\Lambda$CDM ($\mu_k=1$), taking $\Omega^0_{m}=0.3$, it overestimates the growth factor today by $7\%$. For the models of modified gravity that we consider, the precision of~(\ref{growth_factor}) depends on the scale and on the parameters of the theory. This will be discussed in the following subsections, after each model is introduced. 

Given the above analytical approximation for the growth factor, we inverse Fourier transform $4\pi f/k^2$ to obtain the following approximate Green's function for the velocity potential.

\begin{eqnarray}\label{vel_pot_real}
{\mathcal F}(\bar{x}-\bar{x}',a)&\approx&\frac{1}{a^4H}\int^a_0da'a'^3H(a')\cdot\left(\frac{1}{|\bar{x}-\bar{x}'|}\right.\nonumber\\
&&\left.+\frac{3}{2}\Omega_m(a'){\mathcal G}(\bar{x}-\bar{x}',a')\right)
\end{eqnarray}

\noindent In the above, ${\mathcal G}(\bar{x}-\bar{x}',a')$ is the inverse Fourier transform of $4\pi\mu_k(a)/k^2$. Finally, the approximate Green's function for the peculiar velocity is:

\begin{eqnarray}\label{vel__real}
\bar{\mathcal F}_v(\bar{x}-\bar{x}',a)&\approx&\frac{1}{a^4H}\int^a_0da'a'^3H(a')\cdot\left(\frac{\bar{x}'-\bar{x}}{|\bar{x}-\bar{x}'|^3}\right.\nonumber \\
&&\left.+\frac{3}{2}\Omega_m(a')\bar{{\mathcal G}}_v(\bar{x}-\bar{x}',a')\right)\,,
\end{eqnarray}

\noindent where $\bar{{\mathcal G}}_v\equiv\bar{\nabla}{\mathcal G}$. It is also informative to solve eq.~(\ref{Poisson}) for the exact Newtonian potential. One finds:

\begin{equation}\label{phi}
\Phi(\bar{x},a)=-Ga^2\int d^3x'\delta\rho(\bar{x}'){\mathcal G}(\bar{x}-\bar{x}',a)\,.
\end{equation} 

Eq.~(\ref{vel__real}) allows us to understand features of the numerically reconstructed LG velocity dipole for gravity models that introduce time- and scale-dependent modifications to the growth of structure. We focus on models that tune their expansion history to be indistinguishable from $\Lambda$CDM. Therefore, for all the models we consider, $H$ will be parametrized by the $\Lambda$CDM solution with the 5-year WMAP results for the cosmological parameters, i.e. $\Omega_{m0}=0.27$, $H_0=70.5$ $\textrm{km}$ $\textrm{s}^{-1}\textrm{Mpc}^{-1}$, and $\Omega_{\Lambda}=0.73$~\cite{Komatsu:2009}.

In the remainder of this section we specialize to two broad classes of models which encompass alternatives to the $\Lambda$CDM scenario in the context of cosmic acceleration. The first class is scalar-tensor theories  and dark energy models with interactions in the dark sector. The second class is extra dimensional theories of gravity.

\subsection{Scalar-tensor theories and Yukawa-type interactions} \label{Sec:Yukawa}

To describe scalar-tensor theories, e.g. $f(R)$ theories, and models of coupled dark energy, we consider the parametrization introduced by Bertschinger and Zukin~\cite{edbert}. The rescaling of Newton's constant is of the form: 

\begin{equation}\label{BZ}
\mu_k(a)=\frac{1+\alpha_1k^2a^s}{1+\alpha_2k^2a^s}\,.
\end{equation}

\noindent In the above, $\alpha_1, \alpha_2$, and $s$ are free parameters. The parameter $s$ is dimensionless while $\alpha_1$ and $\alpha_2$ have units of length squared. In order to recover the GR solution at early times, we take $s>0$. We take $\alpha_2>0$ so that the coupling is finite, and we take $\alpha_1>0$ so that gravity is attractive. In the specific case of $f(R)$ theories, the ratio $\alpha_1/\alpha_2=4/3$ and $s\sim4$. 

Defining $\bar{\xi}\equiv\bar{x}'-\bar{x}$ and $\xi\equiv|\bar{\xi}|$, the Green functions are: 

\begin{eqnarray}\label{G_BZ}
{\mathcal G}(\xi,a) &=&\frac{1}{\xi}\left[ 1+ \left(\frac{\alpha_1}{\alpha_2}-1\right) e^{-\xi/\sqrt{\alpha_2a^s}}\right]\,,\nonumber\\
\\
\label{Gv_BZ}
\bar{{\mathcal G}}_v(\bar{\xi},a)& =&\frac{\bar{\xi}}{\xi^3}\left[ 1+ \left(\frac{\alpha_1}{\alpha_2}-1\right)\right.\nonumber\\
&&\left.\cdot\left(1+\frac{\xi}{\sqrt{\alpha_2 a^s}}\right)\cdot e^{-\xi/\sqrt{\alpha_2a^s}}\right]\,.
\end{eqnarray}

\noindent The above expression for ${\mathcal G}$, together with eq.~(\ref{phi}), reveals that the Newtonian potential, at late times, contains an additional Yukawa coupling with (comoving) mass $m_{\rm{eff}}=1/\sqrt{\alpha_2a^s}$. The Yukawa correction is expected since alternative theories typically introduce extra degrees of feedom which mediate a fifth force between DM particles.  In the limit that $\xi\, m_{\rm{eff}}\gg1$, the Yukawa coupling is suppressed and we recover GR. The effective coupling, $\alpha_1/\alpha_2-1$, is positive when the fifth force mediator is a scalar field, which is the case for scalar-tensor theories and coupled quintessence. However, it can be negative if the force is mediated by a vector field or a phantom scalar field~\cite{Amendola:2004qb}. Astrophysical and cosmological measurements set combined constraints on the lengthscale and coupling of the fifth force; for scales $100 \,{\rm kpc} \lesssim \sqrt{\alpha_2}\lesssim 10\, {\rm Mpc}$, the intracluster gas distribution~\cite{Gradwohl:1992ue} and cosmological data~\cite{Bean:2008ac} give $0.5\lesssim \alpha_1/\alpha_2\lesssim 2.3$.

We quantify the error of our analytical results by comparing the approximate Green's function (eq.~\ref{vel__real}) with the exact Green's function defined in eq.~(\ref{vel__real_1}). The error depends on the model parameters and the scale $\xi$; it can be traced to how closely $f\sim1$ throughout the expansion history. On large scales, $\mu_k\sim1$ over most of the expansion history, and the analytical Green's functions overestimate the exact solution by $\sim7$\%. On small scales, the errors are smaller (larger) than 7$\%$ for $\alpha_1/\alpha_2>1$ ($<1$) since the growth factor is larger (smaller) than the GR solution. For $\alpha_1/\alpha_2<1$, the error becomes larger on small scales for larger $\sqrt{\alpha_2}$ and smaller $s$, since these all contribute to smaller growth factors. The error for $\sqrt{\alpha_2}$ = 100 Mpc, $\alpha_1/\alpha_2=0.5$, and $s=2$ at $\xi = $10 Mpc is $\sim30$\% --- a factor of 3 larger than typical errors on mass and distance estimates. Hence, the analytical results are accurate to $\sim7\%$ on scales $5 < \xi < 150$ Mpc for $\alpha_1/\alpha_2>1$, but should not be trusted on small scales for $\alpha_1/\alpha_2<1$.      

\subsection{Higher-dimensional gravity} \label{Sec:Degravitation}
Next we consider higher-dimensional extensions of the Dvali-Gabadadze-Porrati (DGP)~\cite{Dvali:2000hr} scenario, within the cascading gravity framework of~\cite{Dvali,Rham}. In this scenario, the matter fields are constrained to a 4-dimensional brane embedded in a higher dimensional bulk of infinite volume, into which the graviton can leak. The  4D gravity carriers are massive gravitons, with a longitudinal mode that mediates a fifth force. The growth of structure for these models has been studied in~\cite{Niayesh,Khoury:2009tk}, where it was found that the Poisson equation is modified by the time- and scale-dependent factor

\begin{equation}\label{mu_Casc}
\mu_k(a)= \frac{1-g(a)}{1+(\frac{a}{kr_c})^{2(1-\alpha)}}\,,
\end{equation} 

\noindent where

\begin{equation}\label{g_Casc}
g(a)=-\left(\frac{D-4}{D-2}\right)\frac{1}{1+2(Hr_c)^{2(1-\alpha)}(1+\frac{H'}{3H})}.
\end{equation}

\noindent $D$ is the number of spacetime dimensions, $0<\alpha<1$ and depends on the number of dimensions, a prime denotes derivates w.r.t. $\ln a$, and $r_c$ is the physical mass scale of the graviton. In~\cite{Niayesh}, it was found that the relevant range for the mass scale is $r_c=300-600\,{\rm Mpc}$. Expression~(\ref{mu_Casc}) describes the combined effects of the massive graviton and the longitudinal mode of the theory. At late times ($Hr_c<1$) and on sub-horizon scales, the growth is enhanced on scales $r<r_c$, by the fifth force mediated by the longitudinal mode. On scales $r>r_c$ the growth is suppressed by the massive graviton.

We specialize to the case $D=6$, which implies $\alpha = 0$ \cite{Niayesh}. We find:

\begin{eqnarray}\label{G_degr}
{\mathcal G}(\xi,a) &=&\frac{1}{\xi}\left(1-g(a)\right) e^{-a\xi/r_c}\,,\nonumber\\
\\
\label{Gv_degr}
\bar{{\mathcal G}}_v(\bar{\xi},a)&=&\frac{\bar{\xi}}{\xi^3}\left(1-g(a)\right)\left(1+\frac{a\xi}{r_c}\right)e^{-a\xi/r_c}\,.\nonumber\\
\end{eqnarray}

\noindent The above expression for ${\mathcal G}$, together with eq.~(\ref{phi}), reveals that the Newtonian potential is modified in amplitude by $(1-g(a))$ and is exponentially cut off on large scales. The effective (comoving) mass for the coupling is given by $m_{\rm{eff}}=a/r_c$. For $\xi \,m_{\rm{eff}}\gg1$, the Newtonian potential exponentially decays. This behavior is different from that scalar tensor theories, where in this limit the $1/\xi$ behavior of GR is recovered on larger scales. This exponential decay decreases the range over which gravity acts. 

As mentioned above, the error in the Green's function, depends on the model parameters and the scale $\xi$. 
The analytical approximation works very well in the relevant range $r_c=300-600\,{\rm Mpc}$. The error in the Green's function for $r_c=300$ Mpc is 3$\%$ at $\xi=5$ Mpc and $7\%$ at $\xi=140$ Mpc. The error gets larger in magnitude for smaller $r_c$; the analytical results should not be trusted for $r_c < 50$ Mpc (where the error at $\xi=150$ Mpc is $\sim 15\%$).    

\subsection{Summary}
As expected, each of the models introduces scale dependent corrections to the Newtonian potential and the peculiar velocity field.  From the Green's functions derived analytically, it is clear how a scale-dependent rescaling of Newton's constant, $\mu_k(a)$, modifies the spatial dependence of the peculiar velocity Green's function. For a given choice of parameters, the contribution of masses at different distances will be modified in a distance-dependent way. As a consequence, the direction of the reconstructed LG dipole will be rotated with respect to the GR result. The amount and direction of this change will depend on the parameters of the models as well as the distribution of masses in the survey. Therefore, it is not straightforward to estimate the amplitude of the change in direction from simple arguments applied to eq.~(\ref{vel__real}).


\section{Reconstructing the Local Group Peculiar Velocity from 2MRS} \label{sec:Reconstruction}

In this section we first provide an overview of the data to be used in the reconstruction (Sec. \ref{Sec:Data} and \ref{Sec:DataMod}), then discuss the method used to compute the peculiar velocity of the LG (Sec. \ref{Sec:Method}).

\subsection{2MRS Groups Catalog} \label{Sec:Data}

The Two Micron All-Sky Survey~\citep[2MASS,][]{Skrutskie:2006} uniformly mapped the entire sky in the J (1.25 $\mu$m), H (1.65 $\mu$m), and K$_s$ (2.16 $\mu$m) bands. By observing in the near-infrared, 2MASS is less susceptible to the severe extinction at low galactic latitudes from which observations at optical wavelengths suffer. Since galaxies spectra peak at $\sim1.6\mu$m, 2MASS maximizes the number of galaxies detected at a specified flux limit, resulting in the most complete all-sky survey performed to date. \textit{IRAS}--selected samples~\citep[e.g.,][]{Strauss:1992a,Fisher:1995,Branchini:1999} also avoid the extinction problems at low galactic latitudes; however, since these samples are based on fluxes in the far-infrared, they are biased toward star-forming galaxies and undersample early-type galaxies; we therefore expect models derived from such samples to underestimate the masses of large clusters.
2MASS makes every effort to obtain an unbiased, uniform sample of extended sources across the entire sky; combined with the ability to detect galaxies at low galactic latitudes, this makes 2MASS an ideal resource to understand the flows in the nearby Universe. In this paper we focus purely on the motion of the LG.

The follow-up 2MASS Redshift Survey~\citep[2MRS,][]{Huchra:2005r,Huchra:2009} is 99.9\% complete to $K_s<11.25$, $b>10^\circ$, and provides redshifts for galaxies selected from the 2MASS extended source catalog.
Efforts to reconstruct the LG dipole from galaxies in 2MASS~\citep{Maller:2003} and 2MRS~\citep{Erdogdu:2006d, Lavaux:2008} have produced results that are both inconsistent with the CMB and depend on the reconstruction method.

The relatively recent work by \citet{Erdogdu:2006d} attempts four different reconstructions for the LG dipole. While the flux-weighted reconstruction in the LG frame suggests a LG dipole $21^\circ \pm 8^\circ$ from the CMB result, when performing the reconstruction in the CMB frame, the result increases to $26^\circ \pm 8^\circ$. Flux-weighted reconstructions necessarily assume an intrinsic mass-to-light ratio, and therefore do not account for variation in dark-matter bias in clusters compared with isolated galaxies. The reconstructions using a number-weighted technique in the same paper suggest a LG velocity more than 35$^\circ$ from the CMB, and since the majority of distances are estimated from redshifts, the appropriate choice for the reference frame becomes unclear.
The more recent reconstruction from 2MRS by \citet{Lavaux:2008} predicts the LG velocity $\sim 50^\circ$ from the CMB direction at $100 h^{-1}$ Mpc, with an estimated error of $22^\circ$ (95\% confidence) from the reconstruction technique.
It is evident that the assumptions built into the method chosen to reconstruct the LG velocity have a noticeable impact on its direction. 

In addition to errors associated with reconstruction methods, there are also errors caused by the finite-size of the survey. 
Assuming a $\Lambda$CDM cosmology, the amplitude of peculiar velocity fluctuations for a sample complete to 140 Mpc\footnote{We choose 140 Mpc since this is the size of the sample we will study in this work.} is approximately 60~\kms. For a 1$\sigma$ fluctuation, given a velocity with respect to the CMB rest frame of $\sim600$~\kms, this implies at most a $6^{\circ}$ discrepancy for a survey limited to the same radius. Therefore the power on large scales neglected due to the finite survey volume is unlikely to be the reason for the discrepancy in direction. 

Since the distribution of dark matter drives the inhomogeneities in the density field, we want to use the best available tracers of dark matter to infer mass. While assuming that baryonic matter traces dark matter provides a reasonable starting point, to further assume that the mass of baryonic matter is proportional to the mass of dark matter ignores any differences between dark matter in isolated galaxies or small groups and massive galaxy clusters. To avoid making this assumption, we will use dynamical mass estimates obtained from galaxy groups: we use the catalog of galaxy groups derived from 2MRS~\citep{Crook:2007,Crook:2008} as the basis for our reconstruction. Two catalogs of groups, constructed using a variable-linking-length percolation algorithm, are presented in the aforementioned work; in this paper, we use groups from the revised 2MRS HDC catalog (constructed with a minimum density contrast $\delta\rho/\rho = 80$); the reason for this choice is that these groups are less prone to interloper contamination than the alternative (LDC) catalog (e.g., \cite{Tovmassian:2009}), and we therefore expect the dynamical mass estimates to be more robust.
This choice, however, comes at the expense of missing some member-galaxies.
The catalog of groups is truncated at a distance of 140 Mpc, where the number of galaxies in the 2MRS $K_s<11.25$ catalog between radii $r$ and $r+\ud r$ falls to half its maximum value. The method we use to reconstruct the LG peculiar velocity is discussed below.

In this work we make an effort to reduce and optimize the assumptions in the reconstruction method. However, the purpose of this work is to understand the effect of modified gravity on the reconstructed LG peculiar velocity. And, while choosing a different method, with different assumptions, will undoubtedly change the direction of the reconstructed dipole, the conclusions presented in this paper are robust, and can be reproduced using the alternative number-weighted reconstruction technique utilized by \cite{Erdogdu:2006d}.

\subsection{Modifications to the catalog} \label{Sec:DataMod}
Since no algorithm to identify groups using only 3 components of phase-space information can guarantee finding virialized groups, we have to understand the limitations of the data. While the algorithm used to construct the 2MRS HDC catalog has been tested and optimized using simulations (e.g, \cite{Ramella:1997,Diaferio:1999}), the groups are still susceptible to interloper contamination.

By examining the sky-distributions and velocity dispersions of the members of the groups with the largest contributions to the LG acceleration, we make the following modifications to the group catalog before continuing:

\begin{enumerate}
\item
A group behind the Virgo cluster at ($\alpha$=12$^h$22$^m$, $\delta$=5$^\circ$) with a LG-centric redshift of $\sim2200$~\kms~has been merged with the Virgo cluster; we manually separate the group by requiring that galaxies assigned to the Virgo cluster inside an ellipse ($a=3.5^\circ$, $b=2^\circ$, with the semimajor axis inclined 30$^\circ$ from North toward West) with redshifts greater than 1850~\kms~are re-assigned to a distinct group.
\item
In the group catalog, the nearby M94 group ($z \sim 300$~\kms) has been merged with both the M106 group ($z \sim 500$~\kms) and the Ursa Major group ($z \sim 1000~\kms$). We manually separate this group as follows: We consider those galaxies inside an ellipse ($a=15^\circ$, $b=10^\circ$, with the semimajor axis inclined 60$^\circ$ from North toward East) centered on ($\alpha$=12$^h$40$^m$, $\delta$=41$^\circ$); those with heliocentric redshifts below 400~\kms~are assigned to the M94 group, and those with redshifts between 400 and 700~\kms~are assigned to the M106 group. The remaining galaxies now appear visually as 3 distinct groups: those inside the ellipse ($a=9^\circ$, $b=6^\circ$, with the semimajor axis inclined 70$^\circ$ from North toward West) centered on ($\alpha$=12$^h$24$^m$, $\delta$=29$^\circ$) are split into two groups, with heliocentric redshifts above and below 900~\kms. Those outside the ellipse comprise the Ursa Major group.
\end{enumerate}
The estimated distances to and masses of the groups are then recomputed using the updated list of members.

\subsection{Method} \label{Sec:Method}

The 2MRS HDC group catalog (see Sec. \ref{Sec:Data} above) is a powerful tool for reconstructing the LG peculiar velocity. However, since we are dealing with a catalog constructed from observational data, we must be careful to avoid its caveats. The catalog was constructed using a variable linking-length percolation algorithm, based on that of~\cite{Huchra:1982}, in an effort to prevent an artificial reduction in the number of groups identified at large distances. The resulting selection function\footnote{We define the selection function as the number of galaxies/groups observed per unit volume as a function of redshift.} of the group catalog will therefore differ from that of 2MRS. In our analysis, we consider only groups with 5 or more members, neglecting smaller groups due to the increased errors in their mass estimates. 

We allow for a linear bias, $b$, between the total matter density and the density in galaxy groups, i.e. $\delta\rho_g/\bar{\rho}_g = b \delta\rho/\bar{\rho}$, where the subscript $g$ denotes galaxy groups. We further assume that mass concentrations are point-like and located at the distances specified in the catalog: $\delta\rho_g = M \delta_D(\vec{r}) - \bar{\rho}_g$. Therefore from eq.(\ref{vel__real_1}) the resulting velocity vector of the LG (which we define to be located at the origin) today is given by
\begin{eqnarray}\label{eq:LGPV}
\vec{v}_\mathrm{LG}=\sum_i \frac{M_i^\mathrm{eff}}{4\pi\bar{\rho}_g b}\bar{{\mathcal F}}_v(\vec{r}_i,a=1)
\end{eqnarray}
where $M_i^\mathrm{eff}$ is the effective mass (see below) of the $i^\mathrm{th}$ group, located at distance $|\vec{r_i}|$~\footnote{Since the groups under consideration are at redshifts below $\sim0.03$, the difference between physical and comoving distances is smaller than the errors on the estimated distances to groups. As such, we do not apply corrections to the distances and masses reported in~\cite{Crook:2007,Crook:2008} to account for the cosmology and variation in scale-factor at the different groups.} and $\bar{{\mathcal F}}_v$ is the numerically computed Green's function defined below eq.~\ref{vel__real_1}.
The normalization of eq.~(\ref{eq:LGPV}) depends on the mean matter density in galaxy groups, $\bar{\rho}_g$, as well as the bias parameter, $b$. For the remainder of this work, we will set $b = 1$, since this is just a scaling parameter that will modify the magnitude of the LG peculiar velocity by a constant factor, but will not affect the direction.

Since we are using data drawn from a magnitude-limited survey, the data are subject to the selection bias of the sample (i.e. fewer galaxies are observed per unit volume at larger redshifts). Since the sample is restricted to small redshifts, we expect that the intrinsic mean mass density does not vary with position, however the mass density we infer by estimating the masses of observed groups will inevitably decrease with redshift. In an effort to reduce the impact of the selection bias on our results, we apply a correction to the estimated group masses such that the effective mass per unit volume does not change with redshift. 

We assume the observed mass density has the functional form
\begin{equation} \label{Eq:ObservedMassDensity}
\rho_g(r) = \bar{\rho}_g \frac{\int^\infty_{(r/r_0)^2} x^{\alpha+1} e^{-x} \ud x}{\int^\infty_0 x^{\alpha+1} e^{-x} \ud x}
\end{equation}
where $\alpha$ and $r_0$ are free parameters. A function of this form is motivated by assuming the luminosity density is independent of spatial position, and the luminosity function has the form of a Schechter function \citep{Schechter:1976}; if the mass density is proportional to the luminosity density, then the observed mass density would follow the form of eq.~(\ref{Eq:ObservedMassDensity}). While we do not require that these assumptions hold true, the functional form satisfies the properties that the observed mass density (i) remains finite at the origin, and (ii) decays at large distances, and therefore it provides a good empirical fit to the data.

The observed mass inside a sphere of radius $r$ is given by
\begin{equation} \label{Eq:CumulativeMass}
M(r) = 4\pi \int_0^r \rho_g(r') \ud r'
\end{equation}
and $\bar{\rho}_g$ is specified by requiring 
\begin{displaymath}
M(D) = M_\mathrm{tot}
\end{displaymath}
where $M_\mathrm{tot}$ is the total mass in observed groups, and $D$ is the extent of the sampled volume. We fit eq.~(\ref{Eq:CumulativeMass}) to $M(r)$ constructed from the catalog of galaxy groups (see Figure \ref{Fig:CumulativeMass}), and obtain $\alpha=0.47$, $r_0=65$ Mpc.
The corresponding $\bar{\rho}_g = 1.24 \times 10^{10}$~\Msun Mpc$^{-3}$. By assigning an effective mass to each group, given by
\begin{equation} \label{Eq:BiasCorrection}
M_i^\mathrm{eff} = \frac{M_i^P}{\rho_g(r_i) / \bar{\rho}_g}
\end{equation}
where the $M_i^P$ are the projected masses~\citep{Bahcall:1981} of the groups taken from the catalog,\footnote{The projected mass estimator assumes that GR is valid on scales of order $\lesssim O(1) Mpc$. The models of modified gravity we consider, display a non-linear mechanism which restores GR on small scales (for which $\delta\rho/\rho\gtrsim 1$), therefore it is reasonable to assume standard GR in the dynamical estimates of masses.} we are assuming that mass we do not observe is correlated with the observed galaxy groups.
\PlaceFigCumulativeMass
%

\section{Results} \label{sec:Results}

Before we discuss the effects of modified gravity on the direction of the dipole, it is helpful to analyze the structure of the reconstructed dipole under GR. The contributions from nearby structures can be understood by reconstructing the dipole using only the mass inside a sphere of radius $r$. The magnitude of the peculiar velocity, plotted as a function of the sphere radius, is shown by the solid line in Figures \ref{Fig:Convergence_Yukawa} and \ref{Fig:Convergence_Degrav}.
As expected, the form of the galactic cartesian components expresses the same features as the reconstruction from the 2MRS galaxy sample \citep{Erdogdu:2006d} inside 10,000~\kms. However, since the galaxies have been grouped, we observe a discretized version of the galaxy-based reconstruction, with slight modifications to the amplitude because we use dynamical mass estimates.

The key features to note are the following. The M94 group is the nearest significant group, at a distance of $\sim4$ Mpc located in Ursa Major at ($l$,$b$) $\approx$ (149$^\circ$, 77$^\circ$); lying near the galactic north pole, this group contributes significantly to the $z$--component of the LG dipole. At $\sim15$ Mpc, the Ursa Major group ($l$,$b$) $\approx$ (145$^\circ$, 67$^\circ$) provides a further attraction toward the galactic pole. The Virgo cluster (2MRS HDC Group \#720) at ($l$,$b$) $\approx$ (285$^\circ$, 73$^\circ$) at a distance of $\sim16$ Mpc provides the largest single contribution to the LG velocity of any group, increasing the LG peculiar velocity in the $z$--direction and decreasing the $y$--component. Centaurus (2MRS HDC Group \#730) at a distance of 57 Mpc at ($l$,$b$) $\approx$ (302$^\circ$, 22$^\circ$) reduces the $y$--component and increases the $x$-- and $z$-- components. The Perseus-Pisces cluster (2MRS HDC \#219) at a distance of 77 Mpc at ($l$,$b$) $\approx$ (150$^\circ$, -13$^\circ$) and the Norma Cluster (A3627, 2MRS HDC \#941) at 72 Mpc, ($l$,$b$) $\approx$ (325$^\circ$, -7$^\circ$) both drive down the $z$--component of the LG peculiar velocity, but the ``tug of war'' in the $x$-- and $y$--directions is clearly evident as the pull of Norma is quickly countered by the opposing Perseus-Pisces.

At larger distances, the volume increases more rapidly with distance and the discrete changes in velocity become less evident. The Coma Cluster (A1656, 2MRS HDC \#745) at 106 Mpc, ($l$,$b$) $\approx$ (59$^\circ$, 88$^\circ$) provides the largest single contribution beyond Perseus-Pisces,\footnote{Note that the Shapley Supercluster at $\sim180$ Mpc is outside the sampled volume.} but the dipole appears to have converged by 120 Mpc,
some 46$^\circ$ from the CMB dipole. 
To assess the propagation of errors from the mass and distance estimates we use a Monte Carlo technique, assuming Gaussian errors of 10\% on the distance to the group, in addition to a Poisson error for the dynamical mass estimate; we find the error on the reconstructed angle is $7^\circ$ ($10^\circ$) at 68\% (90\%) confidence. Furthermore, $N$-body simulations performed to assess the applicability of linear perturbation theory, suggest that the uncertainty in the reconstruction method is $\approx 16^\circ$ ($23^\circ$) at 68\% (90\%) confidence \citep{Crook:2009r}. Combining these error estimates, the discrepancy with the CMB dipole should not exceed 25$^\circ$ at 90\% confidence, and the probability of obtaining an offset of 46$^\circ$ or higher is less than 3\%.
 
It is important to be aware that the conclusions presented below will not change if constructed using an alternative source catalog (e.g. galaxy-based sample) since the general form of the contributions toward the LG peculiar velocity remains the same.

Armed with this insight, we now turn to discuss the effect of modified gravity on the LG peculiar velocity. With a LG dipole under GR directed $\gtrsim 40^\circ$ away from the CMB expectation, it is clear that modified gravity must be able to rotate the direction through an angle of this order of magnitude to provide a solution to the problem.

\subsection{Scalar-tensor theories and Yukawa-type interactions} \label{Sec:Yukawa_Results}
Using the scalar-tensor parametrization shown in eq.~(\ref{BZ}), we numerically calculate the LG peculiar velocity vector for theories that introduce an additional Yukawa interaction. We compute this vector for $s=[2,4]$ and $\alpha_1/\alpha_2$ = [0.5, 1.5, 2.3], as a function of $\sqrt{\alpha_2}$. The values of $\alpha_1/\alpha_2$ are chosen in order to be consistent with current limits~\cite{Gradwohl:1992ue,Bean:2008ac}, as discussed in Sec.~\ref{sec:PecVel}. This range of $s$ is expected for typical Chameleon-type models~\cite{Zhao:2008bn}, and the range of $\sqrt{\alpha_2}$ is chosen so that GR is recovered at early times (i.e. $\mu_k\rightarrow 1$). In Figure \ref{Fig:Ang_Yukawa}, we plot the angular deviation of the reconstructed LG velocity from the GR result for this class of models; the amplitude of the reconstructed velocity, relative to the GR case, is plotted in Figure \ref{Fig:Mag_Yukawa}. From Figure \ref{Fig:Ang_Yukawa}, it is evident that the maximal angular deviation is $\sim3^{\circ}$, which is significantly smaller than the $\sim40^{\circ}$ needed to align the LG velocity with the direction inferred from the CMB dipole. 
\PlaceFigAngYukawa
\PlaceFigMagYukawa

The parameters $s$, $\alpha_1/\alpha_2$, and $\alpha_2$ have different effects on the velocity dipole. The parameter $s$ changes the time at which modified gravity becomes dominant. For larger values of $s$, modified gravity becomes relevant at later times, leading to smaller deviations from GR. Larger $s$ also shifts the features in Figures \ref{Fig:Ang_Yukawa} and \ref{Fig:Mag_Yukawa} to larger values of $\sqrt{\alpha_2}$ since a Yukawa coupling which acts over a larger range compensates for the smaller time duration over which modified gravity is relevant. 
In order to understand the effects of $\alpha_2$ and $\alpha_1/\alpha_2$, let us look at the Green's function~(\ref{Gv_BZ}) derived analytically. As discussed previously, the approximation we have used in deriving~(\ref{Gv_BZ}) works well for values of $\alpha_1/\alpha_2>1$ (and on large scales for $\alpha_1/\alpha_2<1$) and is useful for interpreting the numerical results. 

Consider the deviation caused by modified gravity defined as the fractional difference in the magnitude of the velocity,
\begin{equation} \label{Eq:DeltaV}
\Delta v \equiv \frac{v_{\rm{MG}}-v_{\rm{GR}}}{v_{\rm{GR}}}
\end{equation}
and assume $s$ is fixed. Looking at the approximate Green's function~(eq. \ref{Gv_BZ}), we see that $m_{\rm{eff}}$ sets the transition scale between two asymptotic behaviors: $\Delta v\propto (\alpha_1/\alpha_2-1)$ for $\xi m_{\rm{eff}}\ll 1$ and $\Delta v \sim 0$ for $\xi m_{\rm{eff}}\gg1$. Hence, calculating the LG peculiar velocity effectively becomes a weighted average, with the contribution of a mass to the LG peculiar velocity weighted by $\sim 0.44 (\alpha_1/\alpha_2-1)+1$ for $\xi m_{\rm{eff}} \ll 1$, and by $\sim 1$ for $\xi m_{\rm{eff}} \gg 1$;  both weights are relative to the predicted GR contribution. Since $m_{\rm{eff}}=1/\sqrt{\alpha_2a^s}$, changing $\alpha_2$ changes the scale at which the shift in weighting occurs and changing $\alpha_1/\alpha_2$ changes the weighting itself. 

Note that the sign of $(\alpha_1/\alpha_2-1)$ determines whether the contribution to the LG peculiar velocity from galaxy groups on scales $\xi m_{\rm{eff}}< 1$ is smaller or larger than the GR case. 
In the case of $\alpha_1/\alpha_2 > 1$, the change in direction brings the LG dipole closer to the CMB--implied direction, whereas in the case of $\alpha_1/\alpha_2 < 1$, the shift is away from the CMB dipole. Since the amplitude of the change in angle is significantly less than the error on the reconstruction, we cannot draw any conclusions from this observation.
Since our sample probes $5$ Mpc $<\xi<140$ Mpc, the above analytically computed limits for $\Delta v$ are roughly reproduced in Figure \ref{Fig:Mag_Yukawa} for $\alpha_1/\alpha_2>1$. For $\alpha_1/\alpha_2<1$, the analytic limit for $\xi m_{\rm{eff}}\ll1$ does not match the numerical result since the approximation breaks down.  

Figure \ref{Fig:Ang_Yukawa} shows that the maximal angular deviation from the GR reconstructed dipole occurs for $\sqrt{\alpha_2}$ in the range 20--40 Mpc (for $s=2$) and 30--50 Mpc (for $s=4$), depending on the choice of $\alpha_1/\alpha_2$. This maximum deviation is $\sim$2--3$^{\circ}$, significantly less than the $\sim$40${^\circ}$ required to explain the discrepancy with the CMB. The angular deviation goes to zero at the endpoints of the domain since the dipole direction is insensitive to a constant rescaling of all velocities. The value of $m_{\rm{eff}}$ at which Figure \ref{Fig:Ang_Yukawa} peaks, however, depends on the distribution of galaxies. This is analyzed below by
reconstructing the LG peculiar velocity using only groups inside concentric spheres of increasing radii, which is shown in Figure  \ref{Fig:Convergence_Yukawa}. 
\PlaceFigConvergenceYukawa
This allows us to identify which groups drive the amplitude and direction of the peculiar velocity, and how this changes under specific cases of the Yukawa-type. In what follows, we specialize to the parameter values used for Figure \ref{Fig:Convergence_Yukawa} and discuss the details of the plots. 

The left panel shows the LG peculiar velocity for $\alpha_1/\alpha_2=0.5$ ({\it repulsive} Yukawa), $s=2$, and $\sqrt{\alpha_2} = 30$ and 100 Mpc. For $\sqrt{\alpha_2} = 30$ Mpc, the contribution to the peculiar velocity from the groups inside 30 Mpc (dominated by the M94 and Ursa Major groups, and the Virgo Cluster) is significantly reduced compared to the GR case (since $\alpha_1/\alpha_2 < 1$), and the influence of groups beyond $30\, {\rm Mpc}$ (most notably Centaurus, Perseus-Pisces and Norma) are dampened when compared with GR. 
As such, the relative contribution of Centaurus is less than in GR, so we don't see the negative shift in $y$ or positive shift in $x$ at a distance $\sim 55$ Mpc.
At very large distances, the increased weighting of the groups, and larger volume of the concentric shells, means the contribution to the peculiar velocity from masses within the shell is very sensitive to the distribution of mass, and a departure from GR is expected. However, the amplitude of the contribution to the peculiar velocity from these shells is not sufficient to significantly alter the direction of the dipole.
Since the inner groups dominate the contribution toward the galactic pole, the net result is a small shift in the direction toward the galactic plane, bringing the LG dipole $\sim 2^{\circ}$ farther from the direction expected from the CMB.
For $\sqrt{\alpha_2}=100 \,{\rm Mpc}$, the transition scale is close to the limit of the survey. The resulting peculiar velocity therefore follows the GR result, reduced in amplitude by a constant factor, up to the transition scale. Since the dipole has converged by $\sim120$ Mpc, there is little angular difference from the GR result. Hence, while the amplitude of the LG velocity is reduced, the angle remains consistent with GR to $\sim 1^\circ$.

The right panel shows the case of $\alpha_1/\alpha_2=2.3$ ({\it attractive} Yukawa), $s=2$, and $\sqrt{\alpha_2} = 30$ and 100 Mpc. We see the reverse of the above scenario. Since $\alpha_1/\alpha_2 > 1$, the contribution to the peculiar velocity of groups inside $\sqrt{\alpha_2}$ is amplified. Hence for $\sqrt{\alpha_2}= 30 \,{\rm Mpc}$, the increased contributions from the nearby M94 and Ursa Major groups and the Virgo Cluster increase the $z$--component of the velocity over the GR result. For $\sqrt{\alpha_2}=100 \,{\rm Mpc}$, the contribution from all groups is amplified and the graph follows the GR result up to $\sim100$ Mpc, after which the relative contributions of more distant groups is suppressed, most significantly -- the Coma Cluster -- which dampens the jump in positive $z$. The groups beyond 100 Mpc have large effective masses (due to the correction for the selection bias), and one would expect that changing the weighting of these groups via modified gravity, would have significant impact on the reconstructed velocity. However this effect competes with the isotropy on these scales;
indeed, changing the weighting of an isotropic distribution cannot alter the direction of the dipole. Hence the effect of modified gravity on the contribution toward the LG peculiar velocity from the distant groups is somewhat suppressed.

\subsection{Higher-dimensional gravity} \label{Sec:Degravitation_Results}

Using the parametrization shown in eq.~(\ref{Gv_degr}), we numerically reconstruct the LG peculiar velocity for extra-dimensional theories. Since we have chosen $D=6$, this leaves the crossover scale $r_c$ as the only free parameter. While the relevant range is $300\,{\rm Mpc} < r_c < 600\,{\rm Mpc}$~\cite{Niayesh}, we show the results for $10\,{\rm Mpc} <r_c<600\,{\rm Mpc}$ so the effect of this formulation is evident to the reader. Figure \ref{Fig:Ang_Degrav} shows the angular deviation of the reconstructed velocity from the GR result, plotted as a function of $r_c$; the fractional difference in the magnitude of the velocity is shown in Figure \ref{Fig:Mag_Degrav}.
\PlaceFigAngDegrav
\PlaceFigMagDegrav
\PlaceFigConvergenceDegrav

The parameter $r_c$ causes deviations from GR in two distinct ways. To start with, it sets a transition {\it time} ($Hr_c\lesssim 1$) after which modifications to GR become important. As equation~(\ref{g_Casc}) shows, $g(a)$ evolves from approximately zero at early times, to a value of $\sim-(D-4)/(D-2)$ at late times, (when $Hr_c\ll 1$). Larger $r_c$ causes modified gravity to become relevant at later times, leading to smaller deviations from GR. Moreover, although we restrict ourselves to $D=6$, it is clear that more dimensions will cause larger deviations from GR. 

The parameter $r_c$ also sets a transition {\em{scale}}. At late times ($Hr_c\ll 1$), and on sub-horizon scales, the growth of structure is enhanced for $k>a/r_c$, while it is suppressed for $k<a/r_c$. This behavior is reflected in the approximate Green's function~(\ref{Gv_degr}), where contributions from masses at $\xi\gg r_c/a$ are suppressed. Masses at $\xi\ll r_c/a$ are simply weighted by the factor $1-g(a)$ (integrated over $a$ together with the other standard time-dependent terms).

More specifically for $r_c=300\,{\rm Mpc}$ and $a\xi/r_c\ll1$, according to our analytic results, the contribution to the LG peculiar velocity by a galaxy at $\xi$ is weighted by $\sim 1.21$. For $a\xi/r_c\gg1$, the galaxy's contribution is weighted by $\sim 0.56$. (These weights are found by performing the integral over the scale factor $a$ in eq.~(\ref{vel__real}) and they are defined relative to the predicted GR contribution). In Figure \ref{Fig:Mag_Degrav} the analytic limit for $a\xi/r_c\ll1$ agrees well with the numerical results. We don't expect to reproduce the $a\xi/r_c\gg1$ limit since our sample starts at $\sim5$ Mpc and the analytical results are not reliable for $r_c < 50$ Mpc.      

Figure \ref{Fig:Ang_Degrav} shows that the angular deviation from GR is bigger for smaller $r_c$, with a value of $\sim 6^{\circ}$ at $r_c \sim 20\,{\rm  Mpc}$. However, this small value of $r_c$  is inconsistent with the matter power spectrum and other cosmological observables~\cite{Niayesh}; for values of $r_c$ in the relevant regime $300-600\,{\rm Mpc}$, the angular deviation is less than a degree; therefore these higher-dimensional models cannot explain the observed misalignment angle. 
The angular deviation goes to zero for  $a\xi/r_c\ll1$ since the direction of the LG dipole does not change when the contributions from all the masses are uniformly rescaled. Similarly, we expect the angular deviation to go to zero if the domain were extended to values of $r_c$ such that $a\xi/r_c\gg1$.  

Since the survey is limited to $\xi\lesssim 140\,{\rm Mpc}$, it is straightforward to understand the numerical results for values of $300\,{\rm Mpc} < r_c < 600\,{\rm Mpc}$. As $r_c$ increases, modifications to GR become relevant at later times, leading to the integrated time-dependent correction $(1-g(a))$ having a smaller effect. In addition, 
larger $r_c$ causes more of the masses to be weighted by a common factor, leading to less rotation of the dipole. 

We can again use the components of the peculiar velocity in concentric spheres, as shown in Figure \ref{Fig:Convergence_Degrav}, to understand how different masses contribute to the LG peculiar velocity. The detailed analysis of Figure \ref{Fig:Convergence_Yukawa} in the previous subsection, can be reapplied to Figure \ref{Fig:Convergence_Degrav}. When considering values of $r_c$ close to the distance of the Virgo Cluster, the relative suppression of more distant groups shifts the LG peculiar velocity towards Virgo, and also slightly closer to the CMB--implied direction. For $r_c=300$ Mpc, all galaxy groups satisfy $a\xi/r_c \ll 1$ and so the LG peculiar velocity is simply amplified by $\sim1.21$ with negligible change in the direction.
Note once again that since the amplitude of the change in angle is significantly less than the error on the reconstruction, there is little insight to be gained from the direction of the change.

\subsection{Summary} \label{Sec:Summary_Results}
As expected, the scale-dependent modifications to the Poisson's equation (\ref{Poisson}) rotate the direction of the reconstructed LG peculiar velocity. However, for the broad set of models with realistic parameter choices that we have considered, the change in direction with respect to the standard GR result is of the order of a few degrees, which is significantly smaller than the $\sim 40^{\circ}$ needed to align the LG velocity with the CMB dipole. 

From the analysis it is clear that, if there were a large undetected mass either inside or outside the survey volume, which could rotate the dipole in the right direction, modified gravity models would still predict a direction for the LG peculiar velocity within a few degrees of the GR prediction. In other words, modified gravity would not be significantly more efficient than GR in solving the angular discrepancy.

From Figures \ref{Fig:Mag_Yukawa} and \ref{Fig:Mag_Degrav}, it is clear that for reasonable parameter choices, the magnitude of the LG peculiar velocity differs at most by a factor of $2$ from the GR result. Since there are several uncertainties, like the bias, which are used to determine the magnitude of the LG peculiar velocity, we cannot discount modified gravity on the basis of the magnitude of the reconstructed velocity alone. In other words, we can always find a parameter $b$ of order unity that will allow the peculiar velocity of the LG to equal the magnitude of the peculiar velocity inferred from the CMB dipole.


\section{Conclusions} \label{sec:Summary}

The peculiar velocity of the Local Group, reconstructed from inhomogeneities in the density field, does not agree with the velocity inferred from the dipole anisotropy in the CMB. Using the HDC catalog of galaxy groups~\citep{Crook:2007,Crook:2008} constructed from 2MRS~\citep{Huchra:2005r,Huchra:2009}, 
we investigated, with both numerical and analytical methods, whether allowing for modifications of gravity on large scales, or interactions in the dark sector, can alleviate this discrepancy. Using an approximate expression for the growth factor, we have derived a general analytical formalism that relates peculiar velocities in real space to matter overdensities. This formalism applies to generalized theories of gravity that reproduce GR at early times and predict a modified Poisson equation on sub-horizon scales (eq.~\ref{Poisson}). Moreover, it reproduces numerical results for a broad range of parameter values. 

We have then specialized to two broad classes of models, representing the main approaches to the phenomenon of cosmic acceleration, and reconstructed the corresponding LG peculiar velocity. The models we have chosen have the common feature of introducing scale-dependent modifications to the growth of perturbations, and therefore have the potential of rotating the direction of the peculiar velocity. Specifically, we have considered scalar-tensor theories and models of dark energy coupled to dark matter as well as higher-dimensional gravity~\cite{Dvali,Rham}. For reasonable values of each model's parameters, we find that the direction of the LG peculiar velocity changes by at most $\sim 3^{\circ}$ with respect to the standard GR result. This angle is  much less than the required $\sim 40^{\circ}$ to coincide with the direction of the LG dipole inferred from the CMB.

The inability to reconcile the direction of the LG peculiar velocity with the CMB dipole, under either $\Lambda$CDM or modified gravity, continues to raise numerous questions. The selection-bias of magnitude-limited surveys prevents us from detecting all luminous matter inside the limiting survey volume; however, increasing the magnitude limit in 2MRS does not significantly increase the galaxy count inside 7000~\kms; moreover, since we considered galaxy groups, we expect to have included the most massive structures inside 140 Mpc in our analysis. Furthermore, while previous efforts have relied upon assumptions about the masses of galaxies, we have used masses estimated dynamically from galaxy group members, and therefore have indirectly accounted for dark matter interior to the observed galaxies in clusters.

Even though 2MASS samples galaxies at very low galactic latitudes, it is possible that we have missed a massive contribution to the LG velocity located very close to the galactic plane. While an as-yet undetected contribution to the LG velocity (e.g., a poorly sampled, distant, massive supercluster, or --- as suggested by~\cite{Loeb:2008} --- a nearby galaxy masked by the galactic plane) could reconcile the issue, we find that GR and modified gravity are expected to give directions for the LG velocity that are within a few degrees of each other. A modification of the laws of interaction would therefore not play a key role if this was the case. 

On the other hand, the significant discrepancy between the two dipoles leads us to question the assumptions that are built into the reconstruction. In this and previous work, it is assumed that the luminous matter traces the dark matter. While we have attempted to estimate the cluster masses dynamically (using luminous matter as tracers), we can only estimate interior mass. Our assumption of linear bias implies that the estimated cluster mass is proportional to the dark matter mass, and that this constant of proportionality does not depend on the properties of the cluster, or on redshift, which may not be a valid assumption.

We have shown that changing the scale-dependence of the growth of perturbations does alter the reconstructed direction of the LG dipole, although not in an amount sufficient to explain the discrepancy with the CMB. While we have assumed a linear bias between visible and dark matter, a scale-dependent bias might have a similar effect to that of modified gravity discussed above, but subject to different observational constraints. It is evident that a scale-dependence in the bias parameter will alter the direction of the reconstructed LG velocity; however, from our analysis, we cannot conclude whether or not realistic models of scale-dependence in the bias could account for the angular discrepancy.

An alternative explanation is a limit in our understanding of cosmology.
In a $\Lambda$CDM universe, we expect that the motions of galaxies (and clusters of galaxies) merge with the Hubble flow at $\sim 150$ Mpc scales. However measurements of the power asymmetry in the CMB~\cite{Hoftuft:2009} and of bulk flows~\cite{Kashlinsky,Watkins}, suggest an inhomogeneous distribution of matter on much larger scales. This large-scale power, while inconsistent with $\Lambda$CDM, could explain the discrepancy between dipoles. Indeed, the dipole reconstructed from local inhomogeneities assuming GR, in conjunction with a bulk flow in the direction suggested by both \cite{Kashlinsky} and \cite{Watkins}, can result in a net flow in agreement, within the estimated errors, with the direction observed in the CMB (with an appropriate choice for the bias parameter). Further discussion on this subject, however, is beyond the scope of this work (see~\cite{Crook:2009f} for a detailed analysis).

Peculiar velocities probe the matter density on a wide range of scales, and thus present a unique test of gravity on large sub-horizon scales. While the effect of modified gravity on the LG peculiar velocity is minimal, studying the impact of modified gravity on the local flow-field could potentially act as a more distinguishing test of gravity. Further study on this subject would be highly insightful, but is, once again, beyond the scope of this work.

While in this work we cannot deduce the correct explanation for the discrepancy between the peculiar velocity of the LG induced by inhomogeneities in the density field and the velocity expected from the dipole in the CMB, we have demonstrated that modified theories of gravity alone cannot provide a solution to this problem.

\section*{Acknowledgements}

We thank Ed Bertschinger, John Huchra and Max Tegmark for useful discussions and their feedback on this work.
The work of ACC is supported by the Massachusetts Institute of Technology Whiteman Fellowships and the National Science Foundation under grant AST-0406906. The work of AS and PZ is supported by the National Science Foundation under grant AST-0708501.

\newcommand{\noopsort}[1]{} \newcommand{\singleletter}[1]{#1}


\begin{thebibliography}{50}
\expandafter\ifx\csname natexlab\endcsname\relax\def\natexlab#1{#1}\fi
\expandafter\ifx\csname bibnamefont\endcsname\relax
  \def\bibnamefont#1{#1}\fi
\expandafter\ifx\csname bibfnamefont\endcsname\relax
  \def\bibfnamefont#1{#1}\fi
\expandafter\ifx\csname citenamefont\endcsname\relax
  \def\citenamefont#1{#1}\fi
\expandafter\ifx\csname url\endcsname\relax
  \def\url#1{\texttt{#1}}\fi
\expandafter\ifx\csname urlprefix\endcsname\relax\def\urlprefix{URL }\fi
\providecommand{\bibinfo}[2]{#2}
\providecommand{\eprint}[2][]{\url{#2}}

\bibitem[{\citenamefont{{Penzias} and {Wilson}}(1965)}]{Penzias:1965}
\bibinfo{author}{\bibfnamefont{A.~A.} \bibnamefont{{Penzias}}}
  \bibnamefont{and} \bibinfo{author}{\bibfnamefont{R.~W.}
  \bibnamefont{{Wilson}}}, \bibinfo{journal}{\apj}
  \textbf{\bibinfo{volume}{142}}, \bibinfo{pages}{419} (\bibinfo{year}{1965}).

\bibitem[{\citenamefont{{Corey} and {Wilkinson}}(1976)}]{Corey:1976}
\bibinfo{author}{\bibfnamefont{B.~E.} \bibnamefont{{Corey}}} \bibnamefont{and}
  \bibinfo{author}{\bibfnamefont{D.~T.} \bibnamefont{{Wilkinson}}},
  \bibinfo{journal}{\baas} \textbf{\bibinfo{volume}{8}}, \bibinfo{pages}{351}
  (\bibinfo{year}{1976}).

\bibitem[{\citenamefont{{Fabbri} et~al.}(1980)\citenamefont{{Fabbri}, {Guidi},
  {Melchiorri}, and {Natale}}}]{Fabbri:1980}
\bibinfo{author}{\bibfnamefont{R.}~\bibnamefont{{Fabbri}}},
  \bibinfo{author}{\bibfnamefont{I.}~\bibnamefont{{Guidi}}},
  \bibinfo{author}{\bibfnamefont{F.}~\bibnamefont{{Melchiorri}}},
  \bibnamefont{and} \bibinfo{author}{\bibfnamefont{V.}~\bibnamefont{{Natale}}},
  \bibinfo{journal}{Physical Review Letters} \textbf{\bibinfo{volume}{44}},
  \bibinfo{pages}{1563} (\bibinfo{year}{1980}).

\bibitem[{\citenamefont{{Smoot} et~al.}(1992)}]{Smoot:1992}
\bibinfo{author}{\bibfnamefont{G.~F.} \bibnamefont{{Smoot}}}
  \bibnamefont{et~al.}, \bibinfo{journal}{\apjl}
  \textbf{\bibinfo{volume}{396}}, \bibinfo{pages}{L1} (\bibinfo{year}{1992}).

\bibitem[{\citenamefont{{Bennett} et~al.}(1993)}]{Bennett:1993}
\bibinfo{author}{\bibfnamefont{C.~L.} \bibnamefont{{Bennett}}}
  \bibnamefont{et~al.}, \bibinfo{journal}{Proceedings of the National Academy
  of Science} \textbf{\bibinfo{volume}{90}}, \bibinfo{pages}{4766}
  (\bibinfo{year}{1993}).

\bibitem[{\citenamefont{{Hinshaw} et~al.}(2009)\citenamefont{{Hinshaw},
  {Weiland}, {Hill}, {Odegard}, {Larson}, {Bennett}, {Dunkley}, {Gold},
  {Greason}, {Jarosik} et~al.}}]{Hinshaw:2009}
\bibinfo{author}{\bibfnamefont{G.}~\bibnamefont{{Hinshaw}}},
  \bibinfo{author}{\bibfnamefont{J.~L.} \bibnamefont{{Weiland}}},
  \bibinfo{author}{\bibfnamefont{R.~S.} \bibnamefont{{Hill}}},
  \bibinfo{author}{\bibfnamefont{N.}~\bibnamefont{{Odegard}}},
  \bibinfo{author}{\bibfnamefont{D.}~\bibnamefont{{Larson}}},
  \bibinfo{author}{\bibfnamefont{C.~L.} \bibnamefont{{Bennett}}},
  \bibinfo{author}{\bibfnamefont{J.}~\bibnamefont{{Dunkley}}},
  \bibinfo{author}{\bibfnamefont{B.}~\bibnamefont{{Gold}}},
  \bibinfo{author}{\bibfnamefont{M.~R.} \bibnamefont{{Greason}}},
  \bibinfo{author}{\bibfnamefont{N.}~\bibnamefont{{Jarosik}}},
  \bibnamefont{et~al.}, \bibinfo{journal}{\apjs}
  \textbf{\bibinfo{volume}{180}}, \bibinfo{pages}{225} (\bibinfo{year}{2009}),
  \eprint{0803.0732}.

\bibitem[{\citenamefont{{Yahil} et~al.}(1977)\citenamefont{{Yahil}, {Tammann},
  and {Sandage}}}]{Yahil:1977}
\bibinfo{author}{\bibfnamefont{A.}~\bibnamefont{{Yahil}}},
  \bibinfo{author}{\bibfnamefont{G.~A.} \bibnamefont{{Tammann}}},
  \bibnamefont{and}
  \bibinfo{author}{\bibfnamefont{A.}~\bibnamefont{{Sandage}}},
  \bibinfo{journal}{\apj} \textbf{\bibinfo{volume}{217}}, \bibinfo{pages}{903}
  (\bibinfo{year}{1977}).

\bibitem[{\citenamefont{{Courteau} and {van den Bergh}}(1999)}]{Courteau:1999}
\bibinfo{author}{\bibfnamefont{S.}~\bibnamefont{{Courteau}}} \bibnamefont{and}
  \bibinfo{author}{\bibfnamefont{S.}~\bibnamefont{{van den Bergh}}},
  \bibinfo{journal}{\aj} \textbf{\bibinfo{volume}{118}}, \bibinfo{pages}{337}
  (\bibinfo{year}{1999}).

\bibitem[{\citenamefont{{Maller} et~al.}(2003)\citenamefont{{Maller},
  {McIntosh}, {Katz}, and {Weinberg}}}]{Maller:2003}
\bibinfo{author}{\bibfnamefont{A.~H.} \bibnamefont{{Maller}}},
  \bibinfo{author}{\bibfnamefont{D.~H.} \bibnamefont{{McIntosh}}},
  \bibinfo{author}{\bibfnamefont{N.}~\bibnamefont{{Katz}}}, \bibnamefont{and}
  \bibinfo{author}{\bibfnamefont{M.~D.} \bibnamefont{{Weinberg}}},
  \bibinfo{journal}{\apjl} \textbf{\bibinfo{volume}{598}}, \bibinfo{pages}{L1}
  (\bibinfo{year}{2003}), \eprint{astro-ph/0303592}.

\bibitem[{\citenamefont{{Erdo{\u g}du} et~al.}(2006)\citenamefont{{Erdo{\u
  g}du}, {Huchra}, {Lahav}, {Colless}, {Cutri}, {Falco}, {George}, {Jarrett},
  {Jones}, {Kochanek} et~al.}}]{Erdogdu:2006d}
\bibinfo{author}{\bibfnamefont{P.}~\bibnamefont{{Erdo{\u g}du}}},
  \bibinfo{author}{\bibfnamefont{J.~P.} \bibnamefont{{Huchra}}},
  \bibinfo{author}{\bibfnamefont{O.}~\bibnamefont{{Lahav}}},
  \bibinfo{author}{\bibfnamefont{M.}~\bibnamefont{{Colless}}},
  \bibinfo{author}{\bibfnamefont{R.~M.} \bibnamefont{{Cutri}}},
  \bibinfo{author}{\bibfnamefont{E.}~\bibnamefont{{Falco}}},
  \bibinfo{author}{\bibfnamefont{T.}~\bibnamefont{{George}}},
  \bibinfo{author}{\bibfnamefont{T.}~\bibnamefont{{Jarrett}}},
  \bibinfo{author}{\bibfnamefont{D.~H.} \bibnamefont{{Jones}}},
  \bibinfo{author}{\bibfnamefont{C.~S.} \bibnamefont{{Kochanek}}},
  \bibnamefont{et~al.}, \bibinfo{journal}{\mnras}
  \textbf{\bibinfo{volume}{368}}, \bibinfo{pages}{1515} (\bibinfo{year}{2006}),
  \eprint{arXiv:astro-ph/0507166}.

\bibitem[{\citenamefont{{Webster} et~al.}(1997)\citenamefont{{Webster},
  {Lahav}, and {Fisher}}}]{Webster:1997}
\bibinfo{author}{\bibfnamefont{M.}~\bibnamefont{{Webster}}},
  \bibinfo{author}{\bibfnamefont{O.}~\bibnamefont{{Lahav}}}, \bibnamefont{and}
  \bibinfo{author}{\bibfnamefont{K.}~\bibnamefont{{Fisher}}},
  \bibinfo{journal}{\mnras} \textbf{\bibinfo{volume}{287}},
  \bibinfo{pages}{425} (\bibinfo{year}{1997}), \eprint{arXiv:astro-ph/9608021}.

\bibitem[{\citenamefont{{Rowan-Robinson}
  et~al.}(2000)\citenamefont{{Rowan-Robinson}, {Sharpe}, {Oliver}, {Keeble},
  {Canavezes}, {Saunders}, {Taylor}, {Valentine}, {Frenk}, {Efstathiou}
  et~al.}}]{RowanRobinson:2000}
\bibinfo{author}{\bibfnamefont{M.}~\bibnamefont{{Rowan-Robinson}}},
  \bibinfo{author}{\bibfnamefont{J.}~\bibnamefont{{Sharpe}}},
  \bibinfo{author}{\bibfnamefont{S.~J.} \bibnamefont{{Oliver}}},
  \bibinfo{author}{\bibfnamefont{O.}~\bibnamefont{{Keeble}}},
  \bibinfo{author}{\bibfnamefont{A.}~\bibnamefont{{Canavezes}}},
  \bibinfo{author}{\bibfnamefont{W.}~\bibnamefont{{Saunders}}},
  \bibinfo{author}{\bibfnamefont{A.~N.} \bibnamefont{{Taylor}}},
  \bibinfo{author}{\bibfnamefont{H.}~\bibnamefont{{Valentine}}},
  \bibinfo{author}{\bibfnamefont{C.~S.} \bibnamefont{{Frenk}}},
  \bibinfo{author}{\bibfnamefont{G.~P.} \bibnamefont{{Efstathiou}}},
  \bibnamefont{et~al.}, \bibinfo{journal}{\mnras}
  \textbf{\bibinfo{volume}{314}}, \bibinfo{pages}{375} (\bibinfo{year}{2000}),
  \eprint{arXiv:astro-ph/9912223}.

\bibitem[{\citenamefont{{Lavaux} et~al.}(2008)\citenamefont{{Lavaux}, {Tully},
  {Mohayaee}, and {Colombi}}}]{Lavaux:2008}
\bibinfo{author}{\bibfnamefont{G.}~\bibnamefont{{Lavaux}}},
  \bibinfo{author}{\bibfnamefont{R.~B.} \bibnamefont{{Tully}}},
  \bibinfo{author}{\bibfnamefont{R.}~\bibnamefont{{Mohayaee}}},
  \bibnamefont{and}
  \bibinfo{author}{\bibfnamefont{S.}~\bibnamefont{{Colombi}}},
  \bibinfo{journal}{In press, astro-ph/08103658}  (\bibinfo{year}{2008}),
  \eprint{0810.3658}.

\bibitem[{\citenamefont{Will}(2005)}]{Will}
\bibinfo{author}{\bibfnamefont{C.~M.} \bibnamefont{Will}},
  \bibinfo{journal}{Living Rev. Rel.} \textbf{\bibinfo{volume}{9}},
  \bibinfo{pages}{3} (\bibinfo{year}{2005}), \eprint{gr-qc/0510072}.

\bibitem[{\citenamefont{{Huchra} et~al.}(2005)\citenamefont{{Huchra},
  {Jarrett}, {Skrutskie}, {Cutri}, {Schneider}, {Macri}, {Steining}, {Mader},
  {Martimbeau}, and {George}}}]{Huchra:2005r}
\bibinfo{author}{\bibfnamefont{J.}~\bibnamefont{{Huchra}}},
  \bibinfo{author}{\bibfnamefont{T.}~\bibnamefont{{Jarrett}}},
  \bibinfo{author}{\bibfnamefont{M.}~\bibnamefont{{Skrutskie}}},
  \bibinfo{author}{\bibfnamefont{R.}~\bibnamefont{{Cutri}}},
  \bibinfo{author}{\bibfnamefont{S.}~\bibnamefont{{Schneider}}},
  \bibinfo{author}{\bibfnamefont{L.}~\bibnamefont{{Macri}}},
  \bibinfo{author}{\bibfnamefont{R.}~\bibnamefont{{Steining}}},
  \bibinfo{author}{\bibfnamefont{J.}~\bibnamefont{{Mader}}},
  \bibinfo{author}{\bibfnamefont{N.}~\bibnamefont{{Martimbeau}}},
  \bibnamefont{and} \bibinfo{author}{\bibfnamefont{T.}~\bibnamefont{{George}}},
  in \emph{\bibinfo{booktitle}{Nearby Large-Scale Structures and the Zone of
  Avoidance}} (\bibinfo{year}{2005}), vol. \bibinfo{volume}{329} of
  \emph{\bibinfo{series}{Astronomical Society of the Pacific Conference
  Series}}.

\bibitem[{\citenamefont{{Huchra} et~al.}(2009)}]{Huchra:2009}
\bibinfo{author}{\bibfnamefont{J.~P.} \bibnamefont{{Huchra}}}
  \bibnamefont{et~al.}, \bibinfo{journal}{In preparation}
  (\bibinfo{year}{2009}).

\bibitem[{\citenamefont{{Crook} et~al.}(2007)\citenamefont{{Crook}, {Huchra},
  {Martimbeau}, {Masters}, {Jarrett}, and {Macri}}}]{Crook:2007}
\bibinfo{author}{\bibfnamefont{A.~C.} \bibnamefont{{Crook}}},
  \bibinfo{author}{\bibfnamefont{J.~P.} \bibnamefont{{Huchra}}},
  \bibinfo{author}{\bibfnamefont{N.}~\bibnamefont{{Martimbeau}}},
  \bibinfo{author}{\bibfnamefont{K.~L.} \bibnamefont{{Masters}}},
  \bibinfo{author}{\bibfnamefont{T.}~\bibnamefont{{Jarrett}}},
  \bibnamefont{and} \bibinfo{author}{\bibfnamefont{L.~M.}
  \bibnamefont{{Macri}}}, \bibinfo{journal}{\apj}
  \textbf{\bibinfo{volume}{655}}, \bibinfo{pages}{790} (\bibinfo{year}{2007}),
  \eprint{arXiv:astro-ph/0610732}.

\bibitem[{\citenamefont{{Crook} et~al.}(2008)\citenamefont{{Crook}, {Huchra},
  {Martimbeau}, {Masters}, {Jarrett}, and {Macri}}}]{Crook:2008}
\bibinfo{author}{\bibfnamefont{A.~C.} \bibnamefont{{Crook}}},
  \bibinfo{author}{\bibfnamefont{J.~P.} \bibnamefont{{Huchra}}},
  \bibinfo{author}{\bibfnamefont{N.}~\bibnamefont{{Martimbeau}}},
  \bibinfo{author}{\bibfnamefont{K.~L.} \bibnamefont{{Masters}}},
  \bibinfo{author}{\bibfnamefont{T.}~\bibnamefont{{Jarrett}}},
  \bibnamefont{and} \bibinfo{author}{\bibfnamefont{L.~M.}
  \bibnamefont{{Macri}}}, \bibinfo{journal}{\apj}
  \textbf{\bibinfo{volume}{685}}, \bibinfo{pages}{1320} (\bibinfo{year}{2008}).

\bibitem[{\citenamefont{Dvali et~al.}(2007)\citenamefont{Dvali, Hofmann, and
  Khoury}}]{Dvali}
\bibinfo{author}{\bibfnamefont{G.}~\bibnamefont{Dvali}},
  \bibinfo{author}{\bibfnamefont{S.}~\bibnamefont{Hofmann}}, \bibnamefont{and}
  \bibinfo{author}{\bibfnamefont{J.}~\bibnamefont{Khoury}},
  \bibinfo{journal}{Phys. Rev.} \textbf{\bibinfo{volume}{D76}},
  \bibinfo{pages}{084006} (\bibinfo{year}{2007}), \eprint{hep-th/0703027}.

\bibitem[{\citenamefont{de~Rham et~al.}(2008)\citenamefont{de~Rham, Hofmann,
  Khoury, and Tolley}}]{Rham}
\bibinfo{author}{\bibfnamefont{C.}~\bibnamefont{de~Rham}},
  \bibinfo{author}{\bibfnamefont{S.}~\bibnamefont{Hofmann}},
  \bibinfo{author}{\bibfnamefont{J.}~\bibnamefont{Khoury}}, \bibnamefont{and}
  \bibinfo{author}{\bibfnamefont{A.~J.} \bibnamefont{Tolley}},
  \bibinfo{journal}{JCAP} \textbf{\bibinfo{volume}{0802}}, \bibinfo{pages}{011}
  (\bibinfo{year}{2008}), \eprint{0712.2821}.

\bibitem[{\citenamefont{Bean}(2001)}]{Bean:2001ys}
\bibinfo{author}{\bibfnamefont{R.}~\bibnamefont{Bean}}, \bibinfo{journal}{Phys.
  Rev.} \textbf{\bibinfo{volume}{D64}}, \bibinfo{pages}{123516}
  (\bibinfo{year}{2001}), \eprint{astro-ph/0104464}.

\bibitem[{\citenamefont{Amendola}(2004{\natexlab{a}})}]{Amendola:2003wa}
\bibinfo{author}{\bibfnamefont{L.}~\bibnamefont{Amendola}},
  \bibinfo{journal}{Phys. Rev.} \textbf{\bibinfo{volume}{D69}},
  \bibinfo{pages}{103524} (\bibinfo{year}{2004}{\natexlab{a}}),
  \eprint{astro-ph/0311175}.

\bibitem[{\citenamefont{Pichon and Bernardeau}(1999)}]{Pichon}
\bibinfo{author}{\bibfnamefont{C.}~\bibnamefont{Pichon}} \bibnamefont{and}
  \bibinfo{author}{\bibfnamefont{F.}~\bibnamefont{Bernardeau}}
  (\bibinfo{year}{1999}), \eprint{astro-ph/9902142}.

\bibitem[{\citenamefont{{Wang} and {Steinhardt}}(1998)}]{Wang:1998}
\bibinfo{author}{\bibfnamefont{L.}~\bibnamefont{{Wang}}} \bibnamefont{and}
  \bibinfo{author}{\bibfnamefont{P.~J.} \bibnamefont{{Steinhardt}}},
  \bibinfo{journal}{\apj} \textbf{\bibinfo{volume}{508}}, \bibinfo{pages}{483}
  (\bibinfo{year}{1998}), \eprint{arXiv:astro-ph/9804015}.

\bibitem[{\citenamefont{Linder and Cahn}(2007)}]{Linder}
\bibinfo{author}{\bibfnamefont{E.~V.} \bibnamefont{Linder}} \bibnamefont{and}
  \bibinfo{author}{\bibfnamefont{R.~N.} \bibnamefont{Cahn}},
  \bibinfo{journal}{Astropart. Phys.} \textbf{\bibinfo{volume}{28}},
  \bibinfo{pages}{481} (\bibinfo{year}{2007}), \eprint{astro-ph/0701317}.

\bibitem[{\citenamefont{{Komatsu} et~al.}(2009)\citenamefont{{Komatsu},
  {Dunkley}, {Nolta}, {Bennett}, {Gold}, {Hinshaw}, {Jarosik}, {Larson},
  {Limon}, {Page} et~al.}}]{Komatsu:2009}
\bibinfo{author}{\bibfnamefont{E.}~\bibnamefont{{Komatsu}}},
  \bibinfo{author}{\bibfnamefont{J.}~\bibnamefont{{Dunkley}}},
  \bibinfo{author}{\bibfnamefont{M.~R.} \bibnamefont{{Nolta}}},
  \bibinfo{author}{\bibfnamefont{C.~L.} \bibnamefont{{Bennett}}},
  \bibinfo{author}{\bibfnamefont{B.}~\bibnamefont{{Gold}}},
  \bibinfo{author}{\bibfnamefont{G.}~\bibnamefont{{Hinshaw}}},
  \bibinfo{author}{\bibfnamefont{N.}~\bibnamefont{{Jarosik}}},
  \bibinfo{author}{\bibfnamefont{D.}~\bibnamefont{{Larson}}},
  \bibinfo{author}{\bibfnamefont{M.}~\bibnamefont{{Limon}}},
  \bibinfo{author}{\bibfnamefont{L.}~\bibnamefont{{Page}}},
  \bibnamefont{et~al.}, \bibinfo{journal}{\apjs}
  \textbf{\bibinfo{volume}{180}}, \bibinfo{pages}{330} (\bibinfo{year}{2009}),
  \eprint{0803.0547}.

\bibitem[{\citenamefont{Bertschinger and Zukin}(2008)}]{edbert}
\bibinfo{author}{\bibfnamefont{E.}~\bibnamefont{Bertschinger}}
  \bibnamefont{and} \bibinfo{author}{\bibfnamefont{P.}~\bibnamefont{Zukin}},
  \bibinfo{journal}{Phys. Rev.} \textbf{\bibinfo{volume}{D78}},
  \bibinfo{pages}{024015} (\bibinfo{year}{2008}), \eprint{0801.2431}.

\bibitem[{\citenamefont{Amendola}(2004{\natexlab{b}})}]{Amendola:2004qb}
\bibinfo{author}{\bibfnamefont{L.}~\bibnamefont{Amendola}},
  \bibinfo{journal}{Phys. Rev. Lett.} \textbf{\bibinfo{volume}{93}},
  \bibinfo{pages}{181102} (\bibinfo{year}{2004}{\natexlab{b}}),
  \eprint{hep-th/0409224}.

\bibitem[{\citenamefont{Gradwohl and Frieman}(1992)}]{Gradwohl:1992ue}
\bibinfo{author}{\bibfnamefont{B.-A.} \bibnamefont{Gradwohl}} \bibnamefont{and}
  \bibinfo{author}{\bibfnamefont{J.~A.} \bibnamefont{Frieman}},
  \bibinfo{journal}{Astrophys. J.} \textbf{\bibinfo{volume}{398}},
  \bibinfo{pages}{407} (\bibinfo{year}{1992}).

\bibitem[{\citenamefont{Bean et~al.}(2008)\citenamefont{Bean, Flanagan, Laszlo,
  and Trodden}}]{Bean:2008ac}
\bibinfo{author}{\bibfnamefont{R.}~\bibnamefont{Bean}},
  \bibinfo{author}{\bibfnamefont{E.~E.} \bibnamefont{Flanagan}},
  \bibinfo{author}{\bibfnamefont{I.}~\bibnamefont{Laszlo}}, \bibnamefont{and}
  \bibinfo{author}{\bibfnamefont{M.}~\bibnamefont{Trodden}},
  \bibinfo{journal}{Phys. Rev.} \textbf{\bibinfo{volume}{D78}},
  \bibinfo{pages}{123514} (\bibinfo{year}{2008}), \eprint{0808.1105}.

\bibitem[{\citenamefont{Dvali et~al.}(2000)\citenamefont{Dvali, Gabadadze, and
  Porrati}}]{Dvali:2000hr}
\bibinfo{author}{\bibfnamefont{G.~R.} \bibnamefont{Dvali}},
  \bibinfo{author}{\bibfnamefont{G.}~\bibnamefont{Gabadadze}},
  \bibnamefont{and} \bibinfo{author}{\bibfnamefont{M.}~\bibnamefont{Porrati}},
  \bibinfo{journal}{Phys. Lett.} \textbf{\bibinfo{volume}{B485}},
  \bibinfo{pages}{208} (\bibinfo{year}{2000}), \eprint{hep-th/0005016}.

\bibitem[{\citenamefont{Afshordi et~al.}(2008)\citenamefont{Afshordi,
  Geshnizjani, and Khoury}}]{Niayesh}
\bibinfo{author}{\bibfnamefont{N.}~\bibnamefont{Afshordi}},
  \bibinfo{author}{\bibfnamefont{G.}~\bibnamefont{Geshnizjani}},
  \bibnamefont{and} \bibinfo{author}{\bibfnamefont{J.}~\bibnamefont{Khoury}}
  (\bibinfo{year}{2008}), \eprint{0812.2244}.

\bibitem[{\citenamefont{Khoury and Wyman}(2009)}]{Khoury:2009tk}
\bibinfo{author}{\bibfnamefont{J.}~\bibnamefont{Khoury}} \bibnamefont{and}
  \bibinfo{author}{\bibfnamefont{M.}~\bibnamefont{Wyman}}
  (\bibinfo{year}{2009}), \eprint{0903.1292}.

\bibitem[{\citenamefont{{Skrutskie} et~al.}(2006)}]{Skrutskie:2006}
\bibinfo{author}{\bibfnamefont{M.~F.} \bibnamefont{{Skrutskie}}}
  \bibnamefont{et~al.}, \bibinfo{journal}{\aj} \textbf{\bibinfo{volume}{131}},
  \bibinfo{pages}{1163} (\bibinfo{year}{2006}).

\bibitem[{\citenamefont{{Strauss} et~al.}(1992)\citenamefont{{Strauss},
  {Davis}, {Yahil}, and {Huchra}}}]{Strauss:1992a}
\bibinfo{author}{\bibfnamefont{M.~A.} \bibnamefont{{Strauss}}},
  \bibinfo{author}{\bibfnamefont{M.}~\bibnamefont{{Davis}}},
  \bibinfo{author}{\bibfnamefont{A.}~\bibnamefont{{Yahil}}}, \bibnamefont{and}
  \bibinfo{author}{\bibfnamefont{J.~P.} \bibnamefont{{Huchra}}},
  \bibinfo{journal}{\apj} \textbf{\bibinfo{volume}{385}}, \bibinfo{pages}{421}
  (\bibinfo{year}{1992}).

\bibitem[{\citenamefont{{Fisher} et~al.}(1995)\citenamefont{{Fisher}, {Huchra},
  {Strauss}, {Davis}, {Yahil}, and {Schlegel}}}]{Fisher:1995}
\bibinfo{author}{\bibfnamefont{K.~B.} \bibnamefont{{Fisher}}},
  \bibinfo{author}{\bibfnamefont{J.~P.} \bibnamefont{{Huchra}}},
  \bibinfo{author}{\bibfnamefont{M.~A.} \bibnamefont{{Strauss}}},
  \bibinfo{author}{\bibfnamefont{M.}~\bibnamefont{{Davis}}},
  \bibinfo{author}{\bibfnamefont{A.}~\bibnamefont{{Yahil}}}, \bibnamefont{and}
  \bibinfo{author}{\bibfnamefont{D.}~\bibnamefont{{Schlegel}}},
  \bibinfo{journal}{\apjs} \textbf{\bibinfo{volume}{100}}, \bibinfo{pages}{69}
  (\bibinfo{year}{1995}).

\bibitem[{\citenamefont{{Branchini} et~al.}(1999)}]{Branchini:1999}
\bibinfo{author}{\bibfnamefont{E.}~\bibnamefont{{Branchini}}}
  \bibnamefont{et~al.}, \bibinfo{journal}{\mnras}
  \textbf{\bibinfo{volume}{308}}, \bibinfo{pages}{1} (\bibinfo{year}{1999}).

\bibitem[{\citenamefont{{Tovmassian} and {Plionis}}(2009)}]{Tovmassian:2009}
\bibinfo{author}{\bibfnamefont{H.}~\bibnamefont{{Tovmassian}}}
  \bibnamefont{and}
  \bibinfo{author}{\bibfnamefont{M.}~\bibnamefont{{Plionis}}},
  \bibinfo{journal}{In press, astro-ph/09022555}  (\bibinfo{year}{2009}),
  \eprint{0902.2555}.

\bibitem[{\citenamefont{{Ramella} et~al.}(1997)\citenamefont{{Ramella},
  {Pisani}, and {Geller}}}]{Ramella:1997}
\bibinfo{author}{\bibfnamefont{M.}~\bibnamefont{{Ramella}}},
  \bibinfo{author}{\bibfnamefont{A.}~\bibnamefont{{Pisani}}}, \bibnamefont{and}
  \bibinfo{author}{\bibfnamefont{M.~J.} \bibnamefont{{Geller}}},
  \bibinfo{journal}{\aj} \textbf{\bibinfo{volume}{113}}, \bibinfo{pages}{483}
  (\bibinfo{year}{1997}).

\bibitem[{\citenamefont{{Diaferio} et~al.}(1999)\citenamefont{{Diaferio},
  {Kauffmann}, {Colberg}, and {White}}}]{Diaferio:1999}
\bibinfo{author}{\bibfnamefont{A.}~\bibnamefont{{Diaferio}}},
  \bibinfo{author}{\bibfnamefont{G.}~\bibnamefont{{Kauffmann}}},
  \bibinfo{author}{\bibfnamefont{J.~M.} \bibnamefont{{Colberg}}},
  \bibnamefont{and} \bibinfo{author}{\bibfnamefont{S.~D.~M.}
  \bibnamefont{{White}}}, \bibinfo{journal}{\mnras}
  \textbf{\bibinfo{volume}{307}}, \bibinfo{pages}{537} (\bibinfo{year}{1999}).

\bibitem[{\citenamefont{{Huchra} and {Geller}}(1982)}]{Huchra:1982}
\bibinfo{author}{\bibfnamefont{J.~P.} \bibnamefont{{Huchra}}} \bibnamefont{and}
  \bibinfo{author}{\bibfnamefont{M.~J.} \bibnamefont{{Geller}}},
  \bibinfo{journal}{\apj} \textbf{\bibinfo{volume}{257}}, \bibinfo{pages}{423}
  (\bibinfo{year}{1982}).

\bibitem[{\citenamefont{{Schechter}}(1976)}]{Schechter:1976}
\bibinfo{author}{\bibfnamefont{P.}~\bibnamefont{{Schechter}}},
  \bibinfo{journal}{\apj} \textbf{\bibinfo{volume}{203}}, \bibinfo{pages}{297}
  (\bibinfo{year}{1976}).

\bibitem[{\citenamefont{{Bahcall} and {Tremaine}}(1981)}]{Bahcall:1981}
\bibinfo{author}{\bibfnamefont{J.~N.} \bibnamefont{{Bahcall}}}
  \bibnamefont{and}
  \bibinfo{author}{\bibfnamefont{S.}~\bibnamefont{{Tremaine}}},
  \bibinfo{journal}{\apj} \textbf{\bibinfo{volume}{244}}, \bibinfo{pages}{805}
  (\bibinfo{year}{1981}).

\bibitem[{\citenamefont{{Crook}}(2009)}]{Crook:2009r}
\bibinfo{author}{\bibfnamefont{A.~C.} \bibnamefont{{Crook}}},
  \bibinfo{journal}{In preparation}  (\bibinfo{year}{2009}).

\bibitem[{\citenamefont{Zhao et~al.}(2008)\citenamefont{Zhao, Pogosian,
  Silvestri, and Zylberberg}}]{Zhao:2008bn}
\bibinfo{author}{\bibfnamefont{G.-B.} \bibnamefont{Zhao}},
  \bibinfo{author}{\bibfnamefont{L.}~\bibnamefont{Pogosian}},
  \bibinfo{author}{\bibfnamefont{A.}~\bibnamefont{Silvestri}},
  \bibnamefont{and}
  \bibinfo{author}{\bibfnamefont{J.}~\bibnamefont{Zylberberg}}
  (\bibinfo{year}{2008}), \eprint{0809.3791}.

\bibitem[{\citenamefont{{Loeb} and {Narayan}}(2008)}]{Loeb:2008}
\bibinfo{author}{\bibfnamefont{A.}~\bibnamefont{{Loeb}}} \bibnamefont{and}
  \bibinfo{author}{\bibfnamefont{R.}~\bibnamefont{{Narayan}}},
  \bibinfo{journal}{\mnras} \textbf{\bibinfo{volume}{386}},
  \bibinfo{pages}{2221} (\bibinfo{year}{2008}), \eprint{0711.3809}.

\bibitem[{\citenamefont{{Hoftuft} et~al.}(2009)\citenamefont{{Hoftuft},
  {Eriksen}, {Banday}, {Gorski}, {Hansen}, and {Lilje}}}]{Hoftuft:2009}
\bibinfo{author}{\bibfnamefont{J.}~\bibnamefont{{Hoftuft}}},
  \bibinfo{author}{\bibfnamefont{H.~K.} \bibnamefont{{Eriksen}}},
  \bibinfo{author}{\bibfnamefont{A.~J.} \bibnamefont{{Banday}}},
  \bibinfo{author}{\bibfnamefont{K.~M.} \bibnamefont{{Gorski}}},
  \bibinfo{author}{\bibfnamefont{F.~K.} \bibnamefont{{Hansen}}},
  \bibnamefont{and} \bibinfo{author}{\bibfnamefont{P.~B.}
  \bibnamefont{{Lilje}}}, \bibinfo{journal}{astro-ph/09031229}
  (\bibinfo{year}{2009}), \eprint{0903.1229}.

\bibitem[{\citenamefont{Kashlinsky et~al.}(2008)\citenamefont{Kashlinsky,
  Atrio-Barandela, Kocevski, and Ebeling}}]{Kashlinsky}
\bibinfo{author}{\bibfnamefont{A.}~\bibnamefont{Kashlinsky}},
  \bibinfo{author}{\bibfnamefont{F.}~\bibnamefont{Atrio-Barandela}},
  \bibinfo{author}{\bibfnamefont{D.}~\bibnamefont{Kocevski}}, \bibnamefont{and}
  \bibinfo{author}{\bibfnamefont{H.}~\bibnamefont{Ebeling}}
  (\bibinfo{year}{2008}), \eprint{0809.3734}.

\bibitem[{\citenamefont{Watkins et~al.}(2008)\citenamefont{Watkins, Feldman,
  and Hudson}}]{Watkins}
\bibinfo{author}{\bibfnamefont{R.}~\bibnamefont{Watkins}},
  \bibinfo{author}{\bibfnamefont{H.~A.} \bibnamefont{Feldman}},
  \bibnamefont{and} \bibinfo{author}{\bibfnamefont{M.~J.} \bibnamefont{Hudson}}
  (\bibinfo{year}{2008}), \eprint{0809.4041}.

\bibitem[{\citenamefont{{Crook} et~al.}(2009)}]{Crook:2009f}
\bibinfo{author}{\bibfnamefont{A.~C.} \bibnamefont{{Crook}}}
  \bibnamefont{et~al.}, \bibinfo{journal}{In preparation}
  (\bibinfo{year}{2009}).

\end{thebibliography}
\end{document}